\documentclass[twocolumn,showpacs,preprintnumbers,amsmath,amssymb,prb,superscriptaddress]
{revtex4}
\usepackage{amsmath,amssymb}
\usepackage{graphicx}% Include figure files
\usepackage{dcolumn}% Align table columns on decimal point
\usepackage{bm}% bold math
\usepackage[dvips]{color}

\def\rnum#1{\expandafter{%
\romannumeral #1}}
\def\Rnum#1{\uppercase\expandafter{%
\romannumeral #1}}

\newcommand{\bol}[1]{\boldsymbol #1}

\begin{document}

%\preprint{APS/123-QED}

\title{Field and temperature dependence of the NMR relaxation rate
in the magnetic quadrupolar liquid phase
of spin-$\frac{\bol 1}{\bol 2}$ frustrated ferromagnetic chains}

\author{Masahiro Sato}
\affiliation{Condensed Matter Theory Laboratory, RIKEN, Wako, Saitama
351-0198, Japan}
\affiliation{Department of Physics and Mathematics, Aoyama Gakuin University, 
Sagamihara, Kanagawa 252-5258, Japan}
\author{Toshiya Hikihara}
\affiliation{Department of Physics, Hokkaido University, Sapporo 060-0810, Japan}
\author{Tsutomu Momoi}
\affiliation{Condensed Matter Theory Laboratory, RIKEN, Wako, Saitama
351-0198, Japan}

%\affiliation{%

%\affiliation{

\date{\today}% It is always \today, today,

\begin{abstract}
It is generally difficult to experimentally distinguish magnetic multipolar
orders in spin systems. Recently, it was proposed that
the temperature dependence of the nuclear magnetic resonance 
relaxation rate $1/T_1$ can involve an indirect, but clear signature of 
the field-induced spin nematic or multipolar Tomonaga-Luttinger (TL) liquid 
phase [Phys.\ Rev.\ B {\bf 79}, 060406(R) (2009)]. In this paper, we evaluate 
accurately the field and temperature dependence of $1/T_1$ in 
spin-$\frac{1}{2}$ frustrated $J_1$-$J_2$ chains
combining field-theoretical techniques with numerical data.
Our results demonstrate that isotherms of $1/T_1$ as a function of
magnetic field also exhibit distinctive non-monotonic behavior in spin nematic
TL liquid, in contrast with the standard TL liquid in
the spin-$\frac{1}{2}$ Heisenberg chain.
The relevance of our results to quasi one-dimensional edge-sharing cuprate 
magnets, such as $\rm LiCuVO_4$, is discussed.
\end{abstract}

% PACS, the Physics and Astronomy Classification Scheme.
%75.10.Jm: Quantized spin models
%75.10.Pq: Spin chain models
%75.50.Ee: Antiferromagnetics
%75.40.Cx: Static properties (order parameter, static susceptibility,
%heat capacities, critical exponents, etc.)
%75.30.Kz: Magnetic phase boundaries (including magnetic transitions,
%metamagnetism, etc.)
%75.40.Gb: Dynamic properties (dynamic susceptibility, spin waves, spin
%diffusion, dynamic scaling, etc.)
%76.60.-k: Nuclear magnetic resonance and relaxation

\maketitle

%%%%%%%%%%%%%%%%%%%%%%%%%%%%%%%%%%%%%%%%%%%%%%%%%%%%%%%%
%%%%%%%%%%%%%%%%%%%%%%%%%%%%%%%%%%%%%%%%%%%%%%%%%%%%%%%%
%%%%%%%%%%%%%%%%%%%%%%%%%%%%%%%%%%%%%%%%%%%%%%%%%%%%%%%%
%%%%%%%%%%%%%%%%%%%%%%%%%%%%%%%%%%%%%%%%%%%%%%%%%%%%%%%%
\section{Introduction}\label{sec:Intro}
One of the current topics in solid-state magnetism is multiple-spin ordering
without any single-spin dipole moment.\cite{Andreev,Wen}
Vector and scalar chiral orders and spin
multipolar orders are typical examples expected to appear in real compounds.
The emergence of vector and scalar chiralities accompanies
the spontaneous breakdown of parity or time-reversal symmetries
and these two could be detected indirectly
from parity- or time-reversal-odd observables and related quantities
(e.g., electric polarization in multiferroics,
asymmetric momentum dependence
of spin structure factors,
Hall conductivity, etc).
On the other hand, spin multipolar orders, for example, spin nematic and spin triatic orders,
in spin-$\frac{1}{2}$ magnets can occur
without breaking spatial symmetry, since they are characterized as condensation
of bound multi-magnons\cite{Chubukov,Momoi0,Momoi1,Momoi2,Kecke,Hikihara1} or
spin-triplet resonating valence bond state.\cite{ShindouM}
The absence of both spin long-range order and lattice symmetry breaking 
makes it difficult to find them in experiments.
Furthermore, no clear experimental proof of the spin multipolar orders
has ever been reported,
because there is no established experimental method of
probing spin multipolar orders.

Thanks to recent theoretical studies,
it has been gradually recognized that magnetic multipolar states
occur in several geometrically frustrated spin-$\frac{1}{2}$ magnets, 
especially, in low dimensions.~\cite{Chubukov,Momoi0,Momoi1,Momoi2,Vekua,
Kecke,Hikihara1,Sudan,Ueda,Nishimoto,Zhitomirsky}
In this paper, we focus
on spin-$\frac{1}{2}$ frustrated chains with ferromagnetic (FM)
nearest-neighboring exchange $J_1<0$ and competing antiferromagnetic (AF)
next-nearest-neighboring exchange $J_2>0$. The Hamiltonian is given by
\begin{eqnarray}
\label{eq:zigzag}
{\cal H}_{J_1-J_2} &=& \sum_{n=1,2} \sum_{j} J_n \bm{S}_j\cdot\bm{S}_{j+n}
-H\sum_{j} S^z_j,
\end{eqnarray}
where $\bm{S}_j$ is the spin-$\frac{1}{2}$ operator on the $j$th site, and
$H$ is the applied magnetic field along the $z$ axis.
Recent studies revealed that this model possesses a series of field-induced 
multipolar Tomonaga-Luttinger (TL) liquid phases in high magnetization 
regime.\cite{Vekua,Kecke,Hikihara1,Sato09,Sudan}
These multipolar TL liquids are interpreted as a hard-core Bose gas of
multi-magnon bound states,\cite{Kecke,Hikihara1} where the number $p$ 
of magnons forming a multi-magnon bound state changes successively 
from $p=2$ to 4 with varying $J_2/|J_1|$.
In the multipolar phase with $p=2$, the quadrupolar (or spin nematic) operator $S_j^\pm S_{j+1}^\pm$
exhibits a quasi long-range order. Similarly, the octupolar (or triatic) operator
$S_j^\pm S_{j+1}^\pm S_{j+2}^\pm$ and
the hexadecapolar operator $S_j^\pm S_{j+1}^\pm S_{j+2}^\pm S_{j+3}^\pm$ show a quasi long-range order,
respectively, in the phases with $p=3$ and $p=4$.
In all of these phases, the longitudinal spin correlation also decays 
algebraically,
while the transverse spin correlation is short-ranged.
These characteristic properties of the multipolar TL liquid phases
are distinct from those in the standard TL liquid
(e.g., spin-$\frac{1}{2}$ Heisenberg chain).
Numerical studies\cite{Hikihara1,Sudan} of the quadrupolar and octupolar 
TL liquid phases
showed that the relevant multipolar correlations are dominant
in the high-magnetization regime, whereas the longitudinal spin-density-wave
(SDW) correlation becomes dominant in the lower-magnetization regime.
We call the high-field multipolar states with $p=2$ and 3 
\emph{quadrupolar and octupolar liquids} and the low-field multipolar states
\emph{SDW$_2$ and SDW$_3$ liquids}, respectively, following 
Ref.~\onlinecite{Hikihara1}.

In these multipolar TL liquids, bulk static quantities, such as entropy,
specific heat, and uniform susceptibility, exhibit qualitatively the same
behaviors as those of standard TL liquids,
and hence it is impossible to distinguish multipolar and 
standard TL liquids from them.
To detect direct evidence of their ordering,
we need to measure multiple-spin order parameter or
relevant multiple-spin correlation functions, which is generally difficult.
Instead, to find any indirect evidence, it is necessary to specify 
an effective method of detecting any signature of spin multipolar 
ordering through usual experiments.
Recently, Ref.~\onlinecite{Sato09} has proposed
an experimental way of detecting a signature of multipolar TL liquids,
showing that
the nuclear magnetic resonance (NMR) relaxation rate $1/T_1$
decreases with lowering temperature $T$ in the multipolar TL liquids near saturation.
This is completely different from the behavior of standard TL liquids in
one-dimensional (1D) AF magnets such as the spin-$\frac{1}{2}$ 
Heisenberg chain,
in which $1/T_1$ always increases with lowering $T$
irrespective of the value of applied magnetic field $H$. However, 
it is generally hard to experimentally approach the vicinity
of the saturation field, especially,
in magnets with strong exchange couplings.
In fact, the saturation field of a $J_1$-$J_2$ magnet
$\rm LiCuVO_4$ is about $40-50$~T,~\cite{Enderle,Hagiwara} and to perform NMR
measurement under such a high field is not an easy task.
An experimentally detecting scheme at low magnetic field is 
therefore desirable.
It was also shown in Ref.~\onlinecite{Sato09} that the momentum dependence
of dynamical structure factors in the multipolar TL liquids
shows a distinct difference from that of the spin-$\frac{1}{2}$ AF 
Heisenberg chain.

In this paper, we reexamine the NMR relaxation rate $1/T_1$
in the spin quadrupolar (nematic) and SDW$_2$ TL liquids,
including both low and high-magnetic-field regimes,
and evaluate its field and temperature dependence quantitatively.
To this end, we combine the field-theoretical (bosonization) approach
with the density-matrix renormalization-group (DMRG) method,
substituting numerical values for nonuniversal parameters in analytic results.
We compare these results with the relaxation rate $1/T_1$ of 
the spin-$\frac{1}{2}$ AF Heisenberg chain. 
It is found that, in addition to
the temperature dependence, the field dependence of $1/T_1$
in the nematic and SDW$_2$ liquids clearly differs from that in
the usual TL liquid of the spin-$\frac{1}{2}$ AF chain.
In the $J_1$-$J_2$ chain,
the relaxation rate $1/T_1$ slowly decreases with increase of magnetic field
(or is almost independent of the field) in the low-field SDW$_2$ regime
and increases suddenly in the high-field nematic regime near saturation,
whereas it increases monotonically, as a function of the field, in the usual TL liquid of the spin-$\frac{1}{2}$ AF chain.
The non-monotonic field dependence of $1/T_1$ is a unique characteristic
of the multipolar TL liquids in spin-$\frac{1}{2}$ $J_1$-$J_2$ chains.
Furthermore, if we tune the direction of the field $H$,
we can eliminate the dominant contribution to $1/T_1$
in the multipolar phases and thereby make $1/T_1$ show
exponential decays at low temperature in both nematic and SDW$_2$ liquids.
This exponential decay is caused by a spin gap in the spin transverse 
excitations, which is also a unique property of the multipolar TL liquids.
Our prediction would be applicable in NMR experiments of quasi
one-dimensional (1D) edge-sharing cuprate
magnets, such as
$\rm LiCuVO_4$ (Refs.~\onlinecite{Enderle,Naito,Hagiwara,Buttgen1}),
$\rm Rb_2Cu_2Mo_3O_{12}$ (Ref.~\onlinecite{Hase}),
$\rm PbCuSO_4(OH)_2$ (Refs.~\onlinecite{Kamieniarz,Baran,Yasui,Wolter}),
$\rm LiCu_2O_2$ (Refs.~\onlinecite{Masuda,Masuda2,Park,Seki}) and
$\rm NaCu_2O_2$ (Ref.~\onlinecite{Drechsler}),
whose magnetic properties are expected to be described by
the spin-$\frac12$ $J_1$-$J_2$ model with FM $J_1$ and AF $J_2$.

The paper is organized as follows.
In Secs.~\ref{sec:NMR} and \ref{sec:Multipolar}, we 
review briefly the theory of the NMR relaxation rate $1/T_1$ in electron 
spin systems and effective theories for multipolar and usual TL liquids.
Particularly, in Sec.~\ref{sec:Multipolar}, 
relying on an established field-theoretical method,
we write down the formula of $1/T_1$ in multipolar liquid phases
of the $J_1$-$J_2$ chain~(\ref{eq:zigzag}) and also that in
TL liquid of the spin-$\frac{1}{2}$ AF Heisenberg chain.
Section~\ref{sec:H-T-dep} contains the main results of the present paper, 
which are quantitative estimates of $1/T_1$
in both quadrupolar and usual TL liquids
as a function of magnetic field and temperature.
Section~\ref{sec:RealMagnets} is devoted to a discussion
of some relevant factors in real magnets
which are neglected in the previous sections.
We consider a broad temperature scale beyond the low-energy effective theory,
the relation between electron-nuclear spin interaction and 
direction of field $H$, and the effects of Dzyaloshinsky-Moriya interaction.
Tuning of the field direction is also discussed.
Finally, in Sec.~\ref{sec:Conclusions}, we summarize our results 
and their relevance to real quasi-1D compounds.

%%%%%%%%%%%%%%%%%%%%%%%%%%%%%%%%%%%%%%%%%%%%%%%%%
%%%%%%%%%%%%%%%%%%%%%%%%%%%%%%%%%%%%%%%%%%%%%%%%%
%%%%%%%%%%%%%%%%%%%%%%%%%%%%%%%%%%%%%%%%%%%%%%%%%
%%%%%%%%%%%%%%%%%%%%%%%%%%%%%%%%%%%%%%%%%%%%%%%%%
\section{NMR relaxation rate}
\label{sec:NMR}

Here, we briefly explain the formula of the NMR relaxation rate $1/T_1$ 
in electron-spin
systems. Provided that the NMR relaxation process is
mainly caused by interaction between electron spin ${\bol S}_j$ and
nuclear spin $\bol I$,
the standard perturbation theory for the interaction evaluates
the relaxation rate $1/T_1$ as follows:~\cite{Slichter,Goto}
\begin{eqnarray}
\label{eq:1/T_1}
1/T_1 &\propto& \sum_k \frac{1}{2}|\tilde A_\perp(k)|^2
\left[{\cal S}^{+-}(k,\omega)+{\cal S}^{-+}(k,\omega)\right]
\nonumber\\
&&+|\tilde A_\parallel(k)|^2{\cal S}^{zz}(k,\omega).
\end{eqnarray}
Here, we have assumed that the electron-spin system is in one dimension. 
The electron-spin dynamical structure factor at finite temperature $T$ 
is defined as
\begin{eqnarray}
\label{eq:DyStFactor}
{\cal S}^{\mu\nu}(k,\omega)&=&\sum_j e^{-ik j}
\int_{-\infty}^\infty dt\,\,e^{i\omega t}
\langle S_j^\mu(t)S_0^\nu(0)\rangle_T,
\end{eqnarray}
where $\langle \cdots \rangle_T$ denotes the thermal average. 
The frequency $\omega$ in Eq.~(\ref{eq:1/T_1}) is a given resonant
value of applied oscillating field. Since its magnitude is
generally much smaller than the energy scale of electron systems, we may
take a limit $\omega/k_BT=\beta\omega\to 0$ in Eq.~(\ref{eq:1/T_1}).
The symbols $\tilde A_{\perp}(k)$ and $\tilde A_{\parallel}(k)$ denote
Fourier components of hyperfine
coupling constants between electron and nuclear spins,
which generally stem from
the dipole-dipole interaction
and an SU(2)-invariant exchange interaction $\bold S_j\cdot\bol I$. 
The longitudinal component $\tilde A_\parallel(k)$ originates only from
the dipole-dipole interaction. In quantum spin systems,
$\tilde A_\parallel(k)$ is usually the same order as $\tilde A_\perp(k)$.

The spatial range of interactions between electron spins and a single
nuclear spin is local, that is, at most the order of
the lattice spacing $a$.
Therefore, the $k$ dependence of $\tilde A_{\perp,\parallel}(k)$ could be
negligible. Under the assumption of such a locality, the hyperfine
coupling term in the Hamiltonian is given by
\begin{eqnarray}
\label{eq:tensor}
{\cal H}_{\rm hf}&=& S_{j=0}^\mu {\cal A}_{\mu\nu}I^\nu,
\end{eqnarray}
where the electron site closest to the nucleus is assumed to be $j=0$
and ${\cal A}_{\mu\nu}$ is the real-space hyperfine coupling tensor.
In this case, $1/T_1$ can be approximated as
\begin{eqnarray}
\label{eq:1/T_1_v2}
1/T_1 &\propto& \frac{1}{2}A^2_\perp(j=0)
\left[{\cal S}^{+-}_{j=0}(\omega)+{\cal S}^{-+}_{j=0}(\omega)\right]
\nonumber\\
&&+A^2_\parallel(j=0){\cal S}^{zz}_{j=0}(\omega).
\end{eqnarray}
Here, hyperfine couplings $A_{\perp}(0)$ and
$A_{\parallel}(0)$ are proportional to
${\cal A}_{xx}+{\cal A}_{yy}$ and
$({\cal A}_{xz}^2+{\cal A}_{yz}^2)^{1/2}$, respectively. 
The local dynamical structure factor
${\cal S}^{\mu\nu}_{j=0}(\omega)$ at finite temperature is represented as
\begin{eqnarray}
\label{eq:LocalDyStFactor}
{\cal S}^{\mu\nu}_{j=0}(\omega)&=&
\int_{-\infty}^\infty dt\,\,e^{i\omega t}
\langle S_0^\mu(t)S_0^\nu(0)\rangle_T.
\end{eqnarray}

%%%%%%%%%%%%%%%%%%%%%%%%%%%%%%%%%%%%%%%%%%%%%%%%%
%%%%%%%%%%%%%%%%%%%%%%%%%%%%%%%%%%%%%%%%%%%%%%%%%
%%%%%%%%%%%%%%%%%%%%%%%%%%%%%%%%%%%%%%%%%%%%%%%%%
%%%%%%%%%%%%%%%%%%%%%%%%%%%%%%%%%%%%%%%%%%%%%%%%%
\section{Multipolar and usual TL liquids}
\label{sec:Multipolar}
In this section, we briefly review the effective theory for multipolar
TL-liquid phases in the $J_1$-$J_2$ spin chain~(\ref{eq:zigzag})
and the well-established TL-liquid theory in the spin-$\frac{1}{2}$ AF
Heisenberg chain.
Using these theories, we can derive analytic forms of dynamical
structure factors and $1/T_1$.

%%%%%%%%%%%%%%%%%%%%%%%%%%%%%%%%%%%%%%
%%%%%%%%%%%%%%%%%%%%%%%%%%%%%%%%%%%%%%
%%%%%%%%%%%%%%%%%%%%%%%%%%%%%%%%%%%%%%
%%%%%%%%%%%%%%%%%%%%%%%%%%%%%%%%%%%%%%
\subsection{Theory for multipolar liquid phases}
\label{sec:Multipolar_1}
Two analytic approaches have been developed for
field-induced multipolar phases in
the model~(\ref{eq:zigzag}): One is a weak-coupling
approach~\cite{Vekua,Hikihara1}
where $|J_1|/J_2 \ll 1$ is assumed and $J_1$ is treated as a
perturbation for two decoupled AF-$J_2$ chains. The other is a
phenomenological bosonization in the vicinity of the
saturation field.~\cite{Kecke,Hikihara1}
The former weak-coupling theory can lead to only the quadrupolar
(nematic) phase, but it enables us to calculate various physical
quantities in principle.
The latter approach can treat all of the multipolar phases
(quadrupolar, octupolar, and hexadecapolar),
while the quantities that can be evaluated are somewhat restricted.
Here, we employ the latter approach, which is sufficient
to calculate $1/T_1$ (the former will be used in Sec.~\ref{sec:RealMagnets}).

In the $J_1$-$J_2$ chain with ferromagnetic $J_1 < 0$,
the excitation mode which destabilizes the fully polarized state at the
saturation field $H_c$
is a multimagnon bound state consisting of $p (\ge 2)$ magnons.
Below the saturation field, the soft bound magnons
proliferate to form TL liquid with dominant transverse multipolar
correlation.
The system is thus described as a hard-core Bose gas of
$p$-magnon bound states.
From a numerical calculation of the $p$-magnon excitation energies,
it has been shown that the soft mode to realize the multipolar TL liquid
is two-, three-, and four-magnon bound states with momentum $k=\pi$
for $-2.669<J_1/J_2<0$, $-3.514<J_1/J_2<-2.720$,
and $-3.764<J_1/J_2<-3.514$, respectively.\cite{Kecke,incommeNematic}

From the picture of the hard-core Bose gas,
we may map multipolar operators and the $z$ component of spin operators 
to operators of bosons in the following forms
\begin{eqnarray}
\label{eq:multipolar_boson}
b_j^\dag &=&(-1)^j S_j^-\cdots S_{j+p-1}^-,
\nonumber\\
n_j\equiv b_j^\dag b_j&=& \frac{1}{p}\left(\frac{1}{2}-S_j^z\right),
\end{eqnarray}
where $b_j$ is an annihilation operator of the hard-core boson,
and we have considered the $p$th-order multipolar liquid states
($p=2, 3,$ and $4$ respectively correspond to quadrupolar,
octupolar, and hexadecapolar liquids).
The staggered factor $(-1)^j$ in the first line comes from
the momentum $k=\pi$ of the soft multimagnon bound states.
%the fact that the bottom of multi-magnon bound-state band is always
%located at $k=\pi$ in the above three parameter regions.
To the 1D hard-core Bose gas, we can apply the standard Abelian
bosonization.~\cite{Haldane,Gia_text}
Its low-energy effective Hamiltonian is represented as
\begin{eqnarray}
\label{eq:eff_multipolar}
{\cal H}_{J_1-J_2}^{\rm eff} &=& \int dx \frac{u}{2}
\left[\kappa^{-1}(\partial_x\Phi)^2+\kappa(\partial_x\Theta)^2\right],
\end{eqnarray}
where $x=ja$ is a spatial coordinate,
$\Phi$ and $\Theta$ are a canonical pair of scalar fields,
$\kappa$ is the TL-liquid parameter, and
$u$ is the elementary excitation velocity. We set
$\kappa\to 1$ here at the saturation limit $H \to H_c$ where
the hard-core boson gas becomes equal to a free fermion gas.
It is numerically shown~\cite{Hikihara1,Sudan}
that the TL-liquid parameter $\kappa$ almost
decreases monotonically with lowering magnetic field $H$
in quadrupolar and octupolar liquid phases.
The velocity $u$ is expected to be of the order of $|J_{1,2}|a$,
except for the saturation limit $H \to H_c$ where $u \to 0$.
We note that the TL-liquid theory is valid
only in the temperature regime of $k_BT \ll u/a$.

The boson operator $b_j$ and the density operator $n_j$ are
expressed in terms of $\Phi$ and $\Theta$:~\cite{Haldane,Gia_text}
\begin{eqnarray}
\label{eq:boson_phase}
b_j^\dag &\approx& e^{i\sqrt{\pi}\Theta}\left[f_0
+f_1\cos(\sqrt{4\pi}\Phi+2\pi\rho j)+\cdots\right],
\nonumber\\
n_j &\approx& \rho+\frac{a}{\sqrt{\pi}}\partial_x\Phi
+g_0 \cos(\sqrt{4\pi}\Phi+2\pi\rho j)+\cdots,
\end{eqnarray}
where $\rho=(1/2-M)/p$ is the mean value of the boson density 
and $M=\langle S_j^z\rangle$ is the magnetization per site.
Factors $f_n$ and $g_n$ are nonuniversal constants.
In the present representation, the scaling dimensions of
$e^{qi\sqrt{4\pi}\Phi}$ and $e^{qi\sqrt{\pi}\Theta}$ are,
respectively, $q^2\kappa$ and $q^2/(4\kappa)$.

From Eqs.~(\ref{eq:multipolar_boson})-(\ref{eq:boson_phase}),
two-point correlation functions of $S_j^z$ and
$p$-th multipolar operator
$M_j^{(p)}=\prod_{n=1}^pS_{j+n-1}^-$ at zero temperature are shown
to asymptotically behave like
\begin{align}
\label{eq:correaltion_multipolar}
\langle S_j^z(\tau)S_0^z(0)\rangle_{0} &\approx M^2
-\frac{p^2\kappa}{4\pi^2}
\left[\left(\frac{a}{w}\right)^2+\left(\frac{a}{\bar w}\right)^2\right]
\nonumber\\
&+\frac{g_0^2}{2}\cos(2\pi\rho j)
\left(\frac{a}{|w|}\right)^{2\kappa}+\cdots,
\nonumber\\
\langle M_j^{(p)}(\tau)M_0^{(p)\dag}(0)\rangle_{0}
&\approx (-1)^jf_0^2\left(\frac{a}{|w|}\right)^{1/(2\kappa)}+\cdots
\end{align}
in the $p$-th multipolar TL-liquid phase, where
$\langle \cdots \rangle_{0}$ denotes the expectation value of observables
at zero temperature. (Incidentally, the $p=1$ case
corresponds to the usual TL liquid in spin-$\frac{1}{2}$ AF chain.
See the following subsection.)
Here, $\tau=it$ is imaginary time, and light-cone coordinates
$w$ and $\bar w$ are defined as $(w,\bar w)=(x+iu\tau,x-iu\tau)$.
Equation~(\ref{eq:correaltion_multipolar}) has been
numerically confirmed,~\cite{Vekua,Hikihara1,Sudan}
and this asymptotic behavior survives down to the phase boundary
between multipolar and lower-field vector chiral phases.
The power-law behavior also indicates that
in the high-field regime (narrowly-defined multipolar regime)
with $\kappa>1/2$, the multipolar correlation is
stronger than the longitudinal spin correlation, while the latter
is dominant in the lower-field SDW regime with $\kappa<1/2$.
Namely, $\kappa=1/2$ is the crossover point between multipolar
and SDW regimes.

In the multipolar TL liquid phase, the transverse spin correlation
decays exponentially,
since a finite energy is necessary to violate
$p$-magnon bound states.
Though we cannot compute the transverse spin correlation function and
the multipolar correlation functions
$\langle M_j^{(p')}(\tau)M_0^{(p')\dag}(0)\rangle_0$ with $p' < p$
within the theory of hard-core Bose
gas of multimagnon bound states,
we can explicitly derive
the form of the transverse spin correlation function
$\langle S_j^\pm(\tau)S_0^\mp(0)\rangle_0$,
which is short ranged,~\cite{Sato09} using the weak-coupling theory
in the quadrupolar-liquid phase ($p=2$).
Furthermore, the exponential decay of the $p'$ ($<p$) magnon correlations
including the transverse spin correlation
has been confirmed numerically for the entire regime of
the quadrupolar and octupolar TL liquid phases.\cite{Hikihara1}
The single-magnon gap can be determined by measuring
transverse dynamical structure factors~\cite{Sato09}
${\cal S}^{\pm\mp}(k,\omega)$, 
for example, in neutron scattering experiment. It would be at most the
order of $|J_1|$ or $J_2$.
This fact of the short-ranged transverse spin correlation
in the multipolar TL liquids plays a key role in the following argument.

The slowly decaying mode in the longitudinal spin correlator
in Eq.~(\ref{eq:correaltion_multipolar}) contributes to the leading temperature
dependence of NMR relaxation rate $1/T_1$.
Through an established method based on conformal field 
theory,~\cite{Gia_text,Giamarchi,Giamarchi2,Giamarchi3,Giamarchi4}
we obtain the local dynamical structure factors at finite temperature from
correlation functions at zero temperature and achieve the formula for $1/T_1$
\begin{eqnarray}
\label{eq:1/T1_multipolar}
1/T_1 &\propto& A_{\parallel:0}^2(0)\frac{p^2\kappa}{\pi}
\left(\frac{a}{u}\right)^2\beta^{-1}
\nonumber\\
&&+A_{\parallel:1}^2(0)\frac{g_0^2}{2}\frac{2a}{u}\cos(\kappa\pi)
B(\kappa,1-2\kappa)
\nonumber\\
&&\times\left(\frac{2\pi a}{\beta u}\right)^{2\kappa-1}+\cdots,
\end{eqnarray}
where $B(x,y)$ is the beta function and
$A_{\parallel:0}(j)=A_{\parallel:1}(j)=A_{\parallel}(j)$. 
The first and second terms on the right-hand side are, respectively, 
contributions from the second and third terms
of $\langle S_j^z(\tau)S_0^z(0)\rangle_{0}$ in
Eq.~(\ref{eq:correaltion_multipolar}).
If we take into account the spatial dependence of $A_\parallel(j)$
instead of the formula~(\ref{eq:1/T_1_v2}),
$A_{\parallel:n}(j)$ generally take different values
(but they are expected to be the same order). 
The transverse spin correlation also contributes to $1/T_1$,
but it must be a thermal activation type $\sim e^{-\Delta/(k_BT)}$,
in which $\Delta$ is proportional to the binding energy of $p$ magnons, 
that is, the single-magnon gap.
Therefore, it can be negligible when temperature is
sufficiently smaller than $\Delta/k_B$.
The absence of the contribution from the transverse spin correlation
is a characteristic feature of the multipolar TL liquids.

The temperature dependence of $1/T_1$ given in Eq.~(\ref{eq:1/T1_multipolar})
is consistent with the previous prediction in Ref.~\onlinecite{Sato09}.
The NMR relaxation rate $1/T_1$ diverges with lowering
temperature in the SDW regime ($\kappa<1/2$), whereas it
decays algebraically with lowering $T$
in the high-field multipolar regime ($\kappa>1/2$).
Thus the temperature dependence changes significantly at the crossover point 
$\kappa=1/2$. This feature contrasts sharply 
with the result of usual TL liquids (see the next subsection).
More detailed properties of $1/T_1$ will be discussed
in Sec.\ \ref{sec:H-T-dep}.

%%%%%%%%%%%%%%%%%%%%%%%%%%%%%%%%%%%%%%
%%%%%%%%%%%%%%%%%%%%%%%%%%%%%%%%%%%%%%
%%%%%%%%%%%%%%%%%%%%%%%%%%%%%%%%%%%%%%
%%%%%%%%%%%%%%%%%%%%%%%%%%%%%%%%%%%%%%
\subsection{TL-liquid theory for the spin-$\frac{\bol 1}{\bol 2}$ AF chain}
\label{sec:Multipolar_2_TL}
Here, we quickly summarize the well-known effective theory for 
the spin-$\frac{1}{2}$ AF Heisenberg chain in magnetic field
to compare with the previously mentioned theory for multipolar liquids.
The Hamiltonian is given by
\begin{eqnarray}
\label{eq:AFchain}
{\cal H}_J &=&  \sum_{j} J \bm{S}_j\cdot\bm{S}_{j+1}
-H\sum_{j} S^z_j,
\end{eqnarray}
where $J>0$ is AF exchange coupling.
As is well known, the low-energy physics of this model belongs to the TL-liquid universality
class from zero field to the saturation field $H_c=2J$.
The low-energy Hamiltonian has the same free-boson form as
Eq.~(\ref{eq:eff_multipolar}):
\begin{eqnarray}
\label{eq:eff_TL}
{\cal H}_{J}^{\rm eff} &=& \int dx \frac{v}{2}
\left[K^{-1}(\partial_x\phi)^2+K(\partial_x\theta)^2\right],
\end{eqnarray}
where $\phi$ and $\theta$ are the pair of scalar fields.
The TL-liquid parameter $K$ and the velocity $v$ are
exactly determined from the Bethe
ansatz.\cite{BogoIK1986,QinFYOA1997,CabraHP1998,Essler}
In this model, $K$ changes monotonically 
from 1/2 to 1 when $H$ increases from zero to $H_c$.
Similarly to the multipolar TL liquids, the value of $v$ is
of order of $Ja$ except for the saturation limit $H \to H_c$ where $v \to 0$.
The TL-liquid theory is valid only for $k_BT \ll v/a$.

Spin operators are bosonized as
\begin{align}
\label{eq:spin_boson}
S_{j}^{z}&\approx M+\frac{a}{\sqrt{\pi}}\partial_{x}\phi
+(-1)^{j}a_{1}\cos(\sqrt{4\pi}\phi+2\pi Mj)+\cdots,
\nonumber\\
S_{j}^{+}&\approx{\rm e}^{{\rm i}\sqrt{\pi}\theta}
\left[b_{0}(-1)^{j}+b_{1}\cos(\sqrt{4\pi}\phi+2\pi Mj)+\cdots\right].
\end{align}
Here $a_1$, $b_0$, and $b_1$ are non-universal coefficients,
whose values are numerically determined as a function of
$H$ (or $M$) in Ref.~\onlinecite{Hikihara2}. 
Equations~(\ref{eq:eff_TL}) and (\ref{eq:spin_boson}) enable us
to estimate $1/T_1$. The result is
\begin{eqnarray}
\label{eq:1/T_1_TL}
1/T_1&\propto&A_{\parallel:0}^2(0)\frac{K}{\pi}
\left(\frac{a}{v}\right)^2\beta^{-1}
\nonumber\\
&&+A_{\parallel:1}^2(0)\frac{a_1^2}{2}\frac{2a}{v}\cos(K\pi)
B(K,1-2K)
\nonumber\\
&&\times\left(\frac{2\pi a}{\beta v}\right)^{2K-1}
\nonumber\\
&&+A_{\perp:1}^2(0)b_0^2\frac{2a}{v}\cos\left(\frac{\pi}{4K}\right)
B\left(\frac{1}{4K},1-\frac{1}{2K}\right)
\nonumber\\
&&\times\left(\frac{2\pi a}{\beta v}\right)^{1/(2K)-1}+\cdots,
\end{eqnarray}
where 
$A_{\parallel:0}(j)=A_{\parallel:1}(j)=A_{\parallel}(j)$ and 
$A_{\perp:1}(j)=A_{\perp}(j)$. 
The first and second terms on the right-hand side originate from
the longitudinal spin correlation, while the third term
originates from the transverse spin correlation.
In contrast to the multipolar liquids, both longitudinal and transverse
spin correlations yield power-law-type functions of temperature.
Equation~(\ref{eq:1/T_1_TL}) shows that either
$A_{\parallel:1}$ or $A_{\perp:1}$ term diverges
at low temperatures regardless of the value of $K$,
except for $K=1/2$.\cite{Sachdev}
This diverging behavior of $1/T_1$ with decreasing temperature 
is a common property of usual TL liquids in 1D magnets, 
and is quite different from the decaying
behavior of $1/T_1$ in the multipolar liquid phases.

%%%%%%%%%%%%%%%%%%%%%%%%%%%%%%%%%%%%%%%%%%%%%%%%%
%%%%%%%%%%%%%%%%%%%%%%%%%%%%%%%%%%%%%%%%%%%%%%%%%
%%%%%%%%%%%%%%%%%%%%%%%%%%%%%%%%%%%%%%%%%%%%%%%%%
%%%%%%%%%%%%%%%%%%%%%%%%%%%%%%%%%%%%%%%%%%%%%%%%%
\section{Field and Temperature Dependence of $\bol 1/\bol T_{\bol 1}$}
\label{sec:H-T-dep}
Utilizing the formulas (\ref{eq:1/T1_multipolar}) and
(\ref{eq:1/T_1_TL}), we examine the magnetic-field and temperature
dependence of $1/T_1$ in both multipolar and usual TL liquid phases.
Among the three multipolar phases, we concentrate on the quadrupolar
TL liquid, which is most widely expanded in the $J_1/J_2$-$H$ space
and is believed to be realized in several real magnets.
We show that $1/T_1$ in the quadrupolar TL liquid exhibits
characteristic features distinct from that in the usual TL liquid.
In Eqs.\ (\ref{eq:1/T1_multipolar}) and (\ref{eq:1/T_1_TL}),
the values of hyperfine constants
$A_{\parallel,\perp:n}(j)$ depend on crystal structure,
the spatial relationship between electron and nuclear spins, and
the direction of applied field $H$.
For simplicity, we assume that all the constants $A_{\parallel,\perp:n}(j)$
are equal throughout this section.

Let us first consider $1/T_1$ of the quadrupolar liquid phase.
We can evaluate the expression of $1/T_1$ given in 
Eq.~(\ref{eq:1/T1_multipolar}),
once the parameters $\kappa$, $u$, and $g_0$ are numerically
estimated.
To this end, we perform numerical calculations of
spin-$\frac12$ $J_1$-$J_2$ chains using the DMRG method.
For $\kappa$ and $g_0$, we compute the equal-time
longitudinal spin correlation function $\langle S_j^z(0)S_{j'}^z(0)\rangle_0$
and the spin polarization $\langle S_j^z(0)\rangle_0$
in the ground state.
We then fit the data to an analytic form obtained by
the bosonization method to estimate $\kappa$ and $g_0$.
(The details of the method have been presented
in Refs.~\onlinecite{Hikihara1}, \onlinecite{Hikihara2}, and 
\onlinecite{Hikihara3}.)
For the velocity $u$, we use uniform magnetic susceptibility.
From the bosonization formula~(\ref{eq:boson_phase}),
the susceptibility is given by
\begin{eqnarray}
\label{eq:susceptibility}
\chi_u&=&\partial \langle S_j^z\rangle/\partial H =p^2 \kappa a/(\pi u)
\end{eqnarray}
in the $p$-th order multipolar phase.
We can thus obtain the value of $u$ from the magnetization curve
(i.e., susceptibility) determined by the DMRG calculation.
We determine the values of $\kappa$, $u$, and $g_0$ at
$M=0.05$, 0.1, 0.15, $\cdots$, 0.45 for several values of $J_1/J_2$.
We have found that the TL-liquid parameter $\kappa$
is basically an increasing function of the field $H$
(see Figs.~14 and 18 of Ref.\ \onlinecite{Hikihara1}),
while $g_0$ has a fairly small field dependence.
As expected, $u/a$ is shown to be of order of the 1D couplings $J_1$ and $J_2$
for intermediate magnetization and approaches zero near the saturation. 
For example, in the case of $J_1/J_2=-1$ ($-2$), 
$u/a\simeq 1.04 J_2$ ($0.49J_2$) at $M=0.25$, while 
$u/a\simeq 0.25 J_2$ ($0.1J_2$) at $M=0.4$.

\begin{figure}%[tth]
\begin{center}
\includegraphics[width=8.6cm]{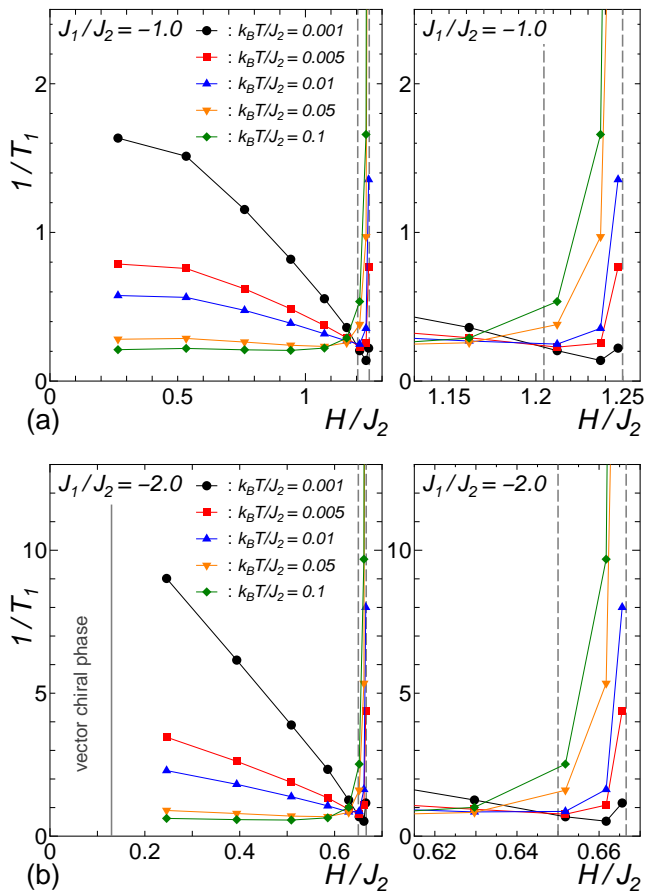}
\end{center}
\caption{(color online)
Field dependence of NMR relaxation rate $1/T_1$ in the magnetic
quadrupolar TL-liquid phase in the $J_1$-$J_2$ spin chain~(\ref{eq:zigzag})
for (a) $J_1/J_2 = -1.0$ and (b) $J_1/J_2 = -2.0$.
Left panels show the results for the whole range of field $0 < H < H_c$
while right panels show the same data in the vicinity of
the saturation field in an enlarged scale.
Solid symbols represent the data obtained from
Eq.\ (\ref{eq:1/T1_multipolar})
with numerical values of $\kappa$, $u$, and $g_0$.
Lines connecting them are guide for the eyes.
The vertical dashed lines represent the saturation field and
the crossover line between the SDW ($\kappa < 1/2$)
and nematic ($\kappa > 1/2$) regimes.
In (b), the boundary to the low-field vector-chiral phase is shown
by vertical solid line.
}
\label{fig:NMR_nematic}
\end{figure}

\begin{figure}%[tth]
\begin{center}
\includegraphics[width=6.5cm]{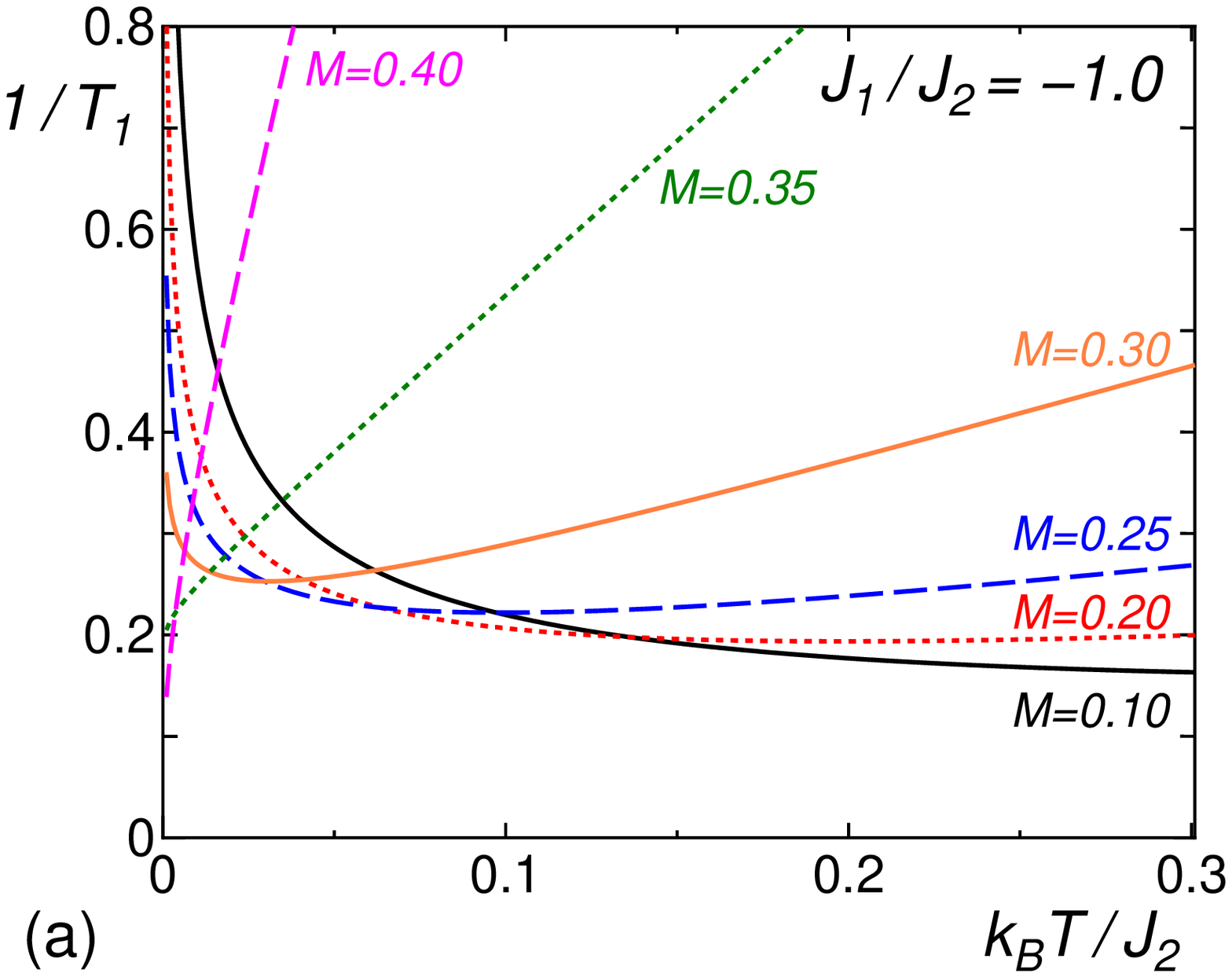}
\includegraphics[width=6.5cm]{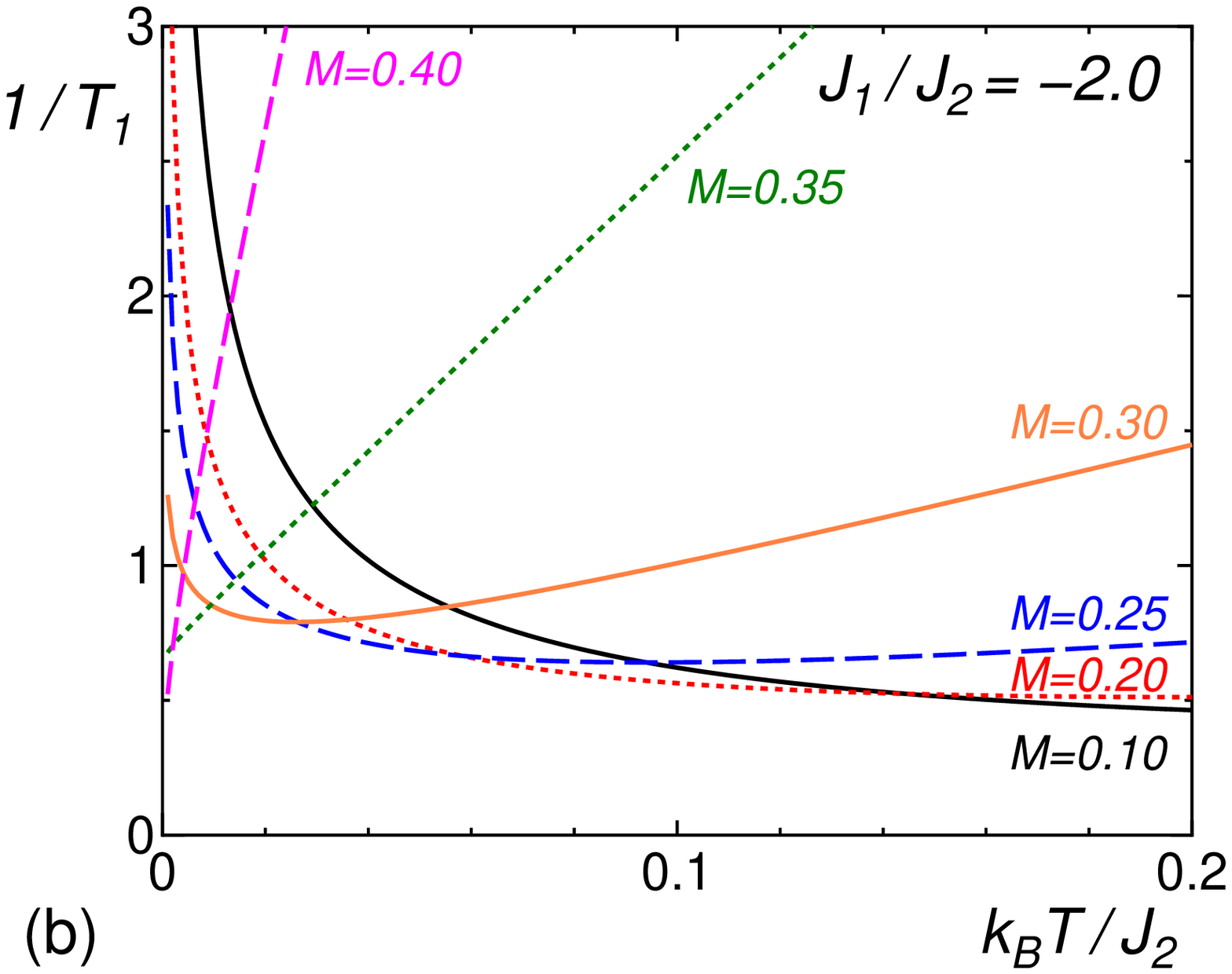}
\end{center}
\caption{(color online)
Temperature dependence of NMR relaxation rate $1/T_1$ in the magnetic
quadrupolar TL-liquid phase in the $J_1$-$J_2$ spin chain~(\ref{eq:zigzag})
for (a) $J_1/J_2 = -1.0$ and (b) $J_1/J_2 = -2.0$ and
several fixed values of magnetization $M$.
}
\label{fig:NMR_nematic_2}
\end{figure}

Figure~\ref{fig:NMR_nematic} shows the magnetic-field dependence
of $1/T_1$ in the $J_1$-$J_2$ spin chain for $J_1/J_2 = -1$ and $-2$
at several values of temperature.
Except for the vicinity of the saturation, 
$1/T_1$ decreases with increasing $H$ (or $M$)
for sufficiently low temperatures $k_BT\ll u/a \sim |J_{1,2}|$.
This is mainly due to the monotonic field dependence of the TL-liquid
parameter $\kappa$, and is a distinct feature of the multipolar TL liquids,
in contrast with that of the spin-$\frac{1}{2}$ AF chain
(see the next paragraph).
This feature becomes less significant as temperature increases, and
it becomes almost independent of the field at relatively high temperatures
$k_BT\sim 0.1 J_2$.
On the other hand, in a narrow region near the saturation field,
$1/T_1$ increases rapidly with $H$ since the velocity $u$ approaches zero. 
We also confirm the previous prediction\cite{Sato09}
on the temperature dependence of $1/T_1$:
$1/T_1$ increases divergently with lowering $T$ in
the low-field SDW regime ($\kappa < 1/2$),
whereas it decreases algebraically in the high-field
quadrupolar regime ($\kappa > 1/2$).
Explicit temperature dependence is
presented in Fig.~\ref{fig:NMR_nematic_2}.
The characteristic behavior in the low-temperature regime
$k_BT\alt 0.1 J_2$ is consistent with the previous statement.

\begin{figure}%[tth]
\begin{center}
\includegraphics[width=6.5cm]{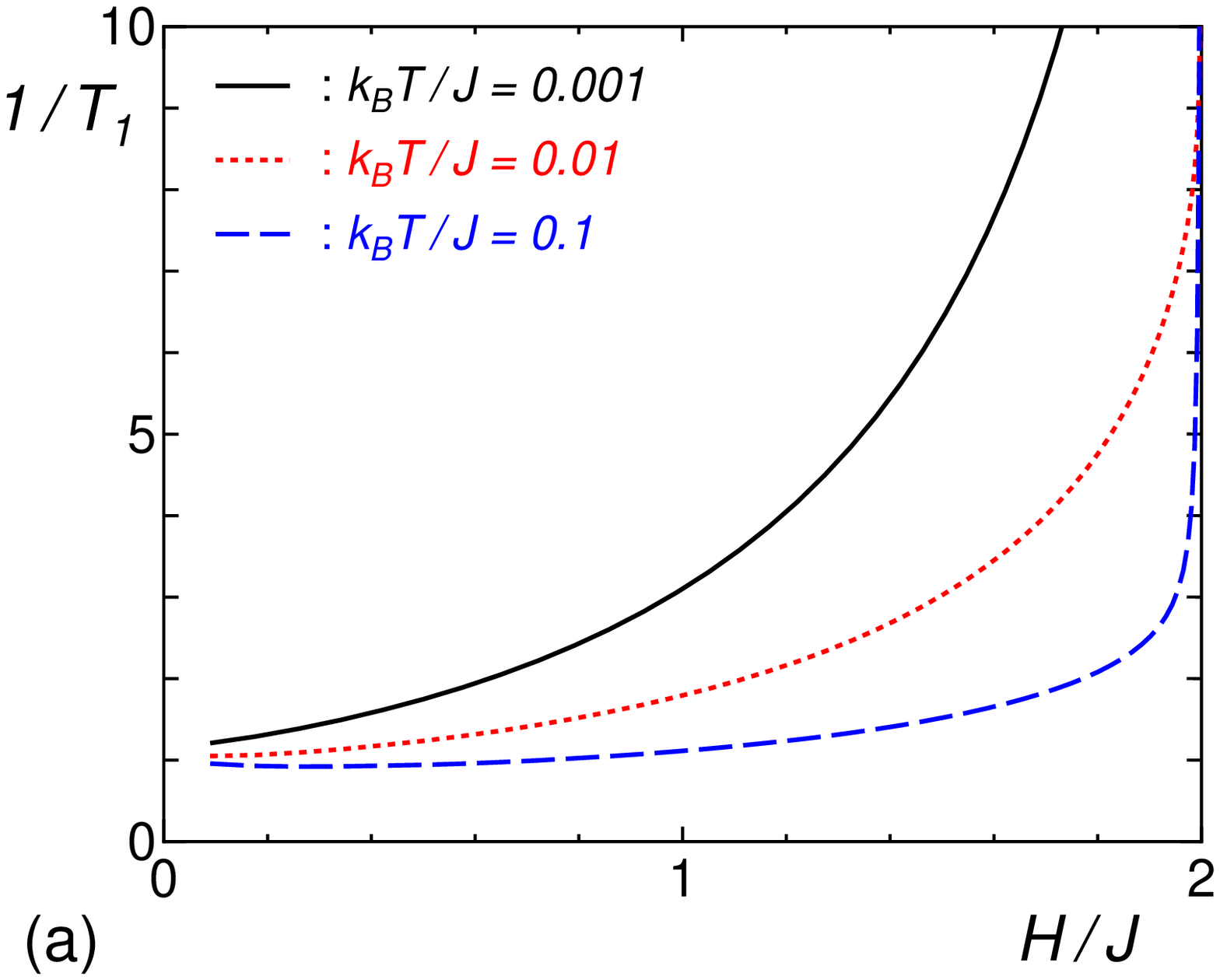}
\includegraphics[width=6.5cm]{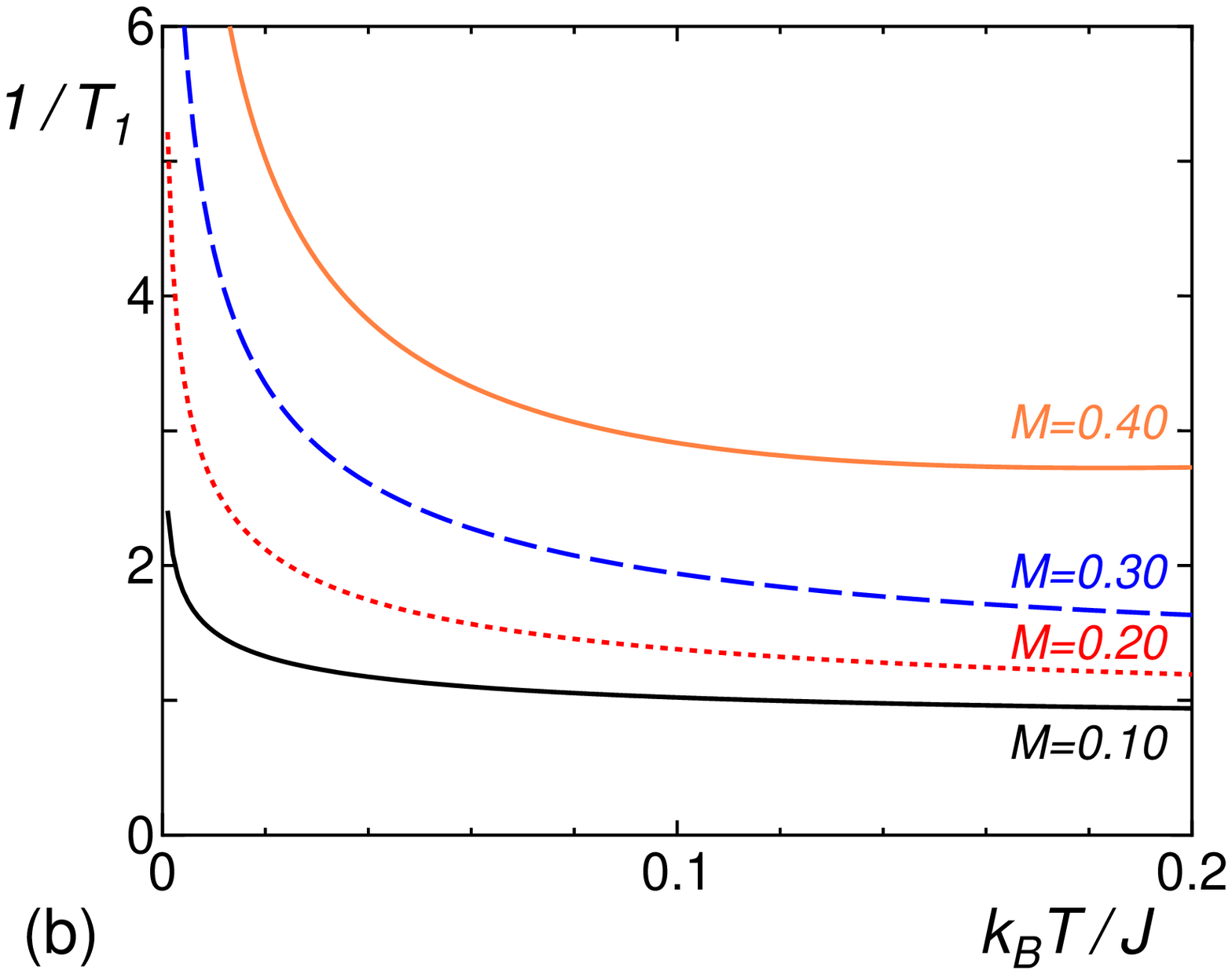}
\end{center}
\caption{(color online)
NMR relaxation rate $1/T_1$ for the standard TL-liquid phase
in the spin-$\frac{1}{2}$ Heisenberg chain~(\ref{eq:AFchain}); 
(a) Field dependences for fixed temperatures and
(b) temperature dependences for fixed magnetization.
}
\label{fig:NMR_usualTL}
\end{figure}

Next, we compare these features of the relaxation rate
in the nematic liquid with those in a usual TL liquid of
the spin-$\frac{1}{2}$ AF Heisenberg chain~(\ref{eq:AFchain}).
Similarly to the nematic case, we can evaluate $1/T_1$ from
the formula~(\ref{eq:1/T_1_TL}) if values for
the parameters $K$, $v$, $a_1$, and $b_0$
are prepared.
As mentioned in Sec.\ \ref{sec:Multipolar_2_TL},
$K$ and $v$ are exactly determined by the Bethe
ansatz\cite{BogoIK1986,QinFYOA1997,CabraHP1998,Essler}
and the coefficients $a_1$ and $b_0$ have
been evaluated numerically.\cite{Hikihara2}
Using these accurate values, we numerically determine $1/T_1$ of
the model~(\ref{eq:AFchain}). The result is shown in
Fig.~\ref{fig:NMR_usualTL}. It indicates that
$1/T_1$ is a monotonically increasing (decreasing) function of
$H$ ($T$) in the TL-liquid phase of the model~(\ref{eq:AFchain})
for sufficiently low temperatures.
These two properties are completely different from those
of the nematic TL-liquid phase shown in Figs.~\ref{fig:NMR_nematic}
and \ref{fig:NMR_nematic_2}. The relaxation rate of real
quasi-1D spin-$\frac{1}{2}$ AF magnets has been
observed in somewhat restricted regime of $T$ and $H$, for example,
in Refs.~\onlinecite{Azevedo,Azevedo2,Groen,Kuhne}.
Their results
seem to agree nicely with our results in Fig.~\ref{fig:NMR_usualTL}.

\begin{figure}%[tth]
\begin{center}
\includegraphics[width=8.6cm]{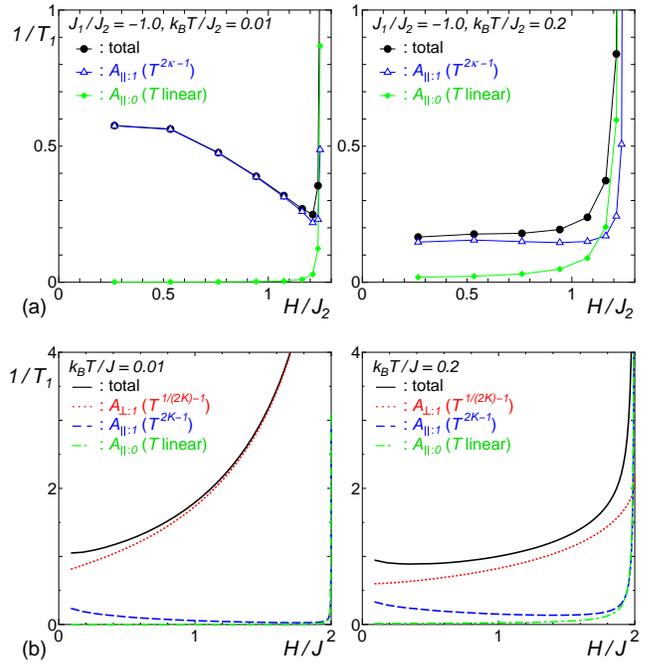}
\end{center}
\caption{(color online)
Field dependence of NMR relaxation rate $1/T_1$ for
(a) the magnetic quadrupolar TL-liquid phase in the $J_1$-$J_2$
spin chain~(\ref{eq:zigzag}) with $J_1/J_2 = -1.0$
and (b) the standard TL-liquid phase
in the spin-$\frac{1}{2}$ Heisenberg chain~(\ref{eq:AFchain}).
Left (right) panels show the data for
$k_BT/J_2 = 0.01$ and $k_BT/J = 0.01$ ($k_BT/J_2 = 0.2$ and $k_BT/J = 0.2$).
In addition to the total value of $1/T_1$,
each term in Eqs.~(\ref{eq:1/T1_multipolar}) and (\ref{eq:1/T_1_TL})
is shown separately.
}
\label{fig:NMR_Each}
\end{figure}

Finally, let us separately estimate each term in the expressions of $1/T_1$,
Eqs.~(\ref{eq:1/T1_multipolar}) and (\ref{eq:1/T_1_TL}).
They are depicted in Fig.~\ref{fig:NMR_Each}.
We find that in the Heisenberg chain~(\ref{eq:AFchain})
the relaxation rate $1/T_1$ is clearly governed by the contribution
from the transverse spin correlation, that is, the $A_{\perp:1}$ term.
In the nematic case, in which the transverse spin fluctuation is gapped,
the $A_{\parallel:1}$ term becomes dominant in a wide
magnetic-field range and is responsible for the behavior of $1/T_1$
decreasing with field $H$. 
This figure definitely demonstrates that characteristic features of
$1/T_1$ in the $J_1$-$J_2$ spin chain originates from the absence
of gapless modes in the transverse spin excitations.

From these results, we conclude
that the NMR relaxation rates $1/T_1$ in the nematic and standard TL-liquid 
phases exhibit quite different behavior in $T$-$H$ space, and 
hence we propose that its $T$ and $H$ dependence can be used to distinguish
the nematic liquid phase from the standard TL liquid.
Similar characteristic behavior of $1/T_1$ is also expected in
the higher-order multipolar liquid phases in the $J_1$-$J_2$
spin chain, since these phases share essentially the same properties of 
spin correlations and the field $H$ dependence of the TL-liquid parameter.
Furthermore, a nematic TL-liquid phase in the SDW regime ($\kappa>1/2$)
is also shown to appear in a wide parameter region of the
$J_1$-$J_2$ chain with AF $J_1$.~\cite{Hikihara3,Okunishi}
This AF-$J_1$ SDW$_2$ phase, where the transverse spin correlation is
short-ranged, is also expected to show a peculiar $H$ dependence of $1/T_1$,
which is distinct from that in the usual TL liquid.

%%%%%%%%%%%%%%%%%%%%%%%%%%%%%%%%%%%%%%%%%%%%%%%%%
%%%%%%%%%%%%%%%%%%%%%%%%%%%%%%%%%%%%%%%%%%%%%%%%%
%%%%%%%%%%%%%%%%%%%%%%%%%%%%%%%%%%%%%%%%%%%%%%%%%
%%%%%%%%%%%%%%%%%%%%%%%%%%%%%%%%%%%%%%%%%%%%%%%%%
\section{Wide Temperature Range, Hyperfine Couplings, and DM interactions}
\label{sec:RealMagnets}
So far we have discussed low-temperature behavior in spin isotropic pure 1D 
systems. In this section, we consider the temperature dependence of $1/T_1$ 
in wide temperature range, taking interchain couplings into account. 
We also discuss the relation between the form of hyperfine coupling tensor 
$\cal A_{\mu\nu}$ and the direction of field $H$, as well as effects of 
Dzyaloshinsky-Moriya (DM) interactions.

%%%%%%%%%%%%%%%%%%%%%%%%%%%%%%%%%%%%%%%%%%%%%%%%%%%%%%%%%%
\subsection{Temperature dependence of $1/T_1$ in a wide temperature range}
\begin{figure}%[tth]
\begin{center}
\includegraphics[width=7cm]{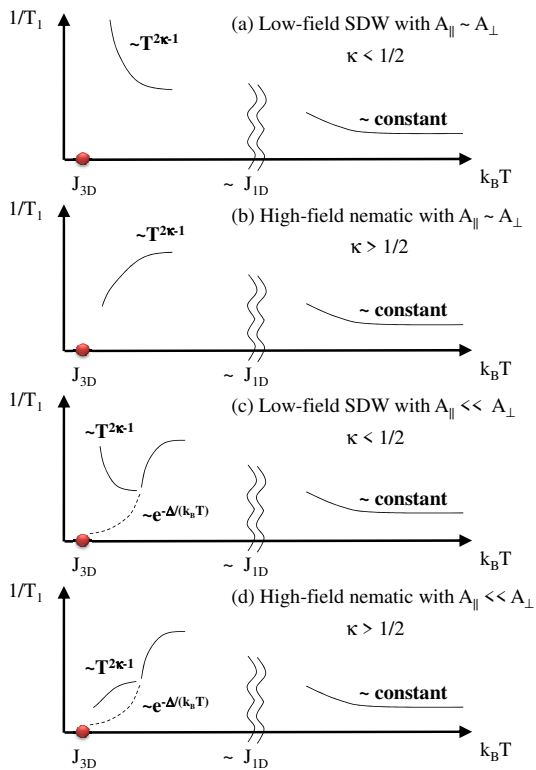}
\end{center}
\caption{(color online) Schematic behavior of $1/T_1$ of the nematic
TL liquid in wide temperature region for several parameter settings.
In the regime $J_{3D} \ll k_BT \ll J_{1D}$, our prediction based on low-energy
effective theories can be applicable.}
\label{fig:several_1/T_1}
\end{figure}

First we consider a wide temperature window including
higher and lower temperatures in which
simple effective Hamiltonians~(\ref{eq:eff_multipolar}) and
(\ref{eq:eff_TL}) are no longer valid for describing the physics of real
quasi-1D magnets.
At sufficiently low temperature, which is lower than the energy scale
of weak 3D interchain couplings $J_{3D}$,
long-range-ordered phases usually emerge due to 3D couplings
and hence 1D effective theories cannot be applicable.
In the case of the SDW regime,
$1/T_1$ exhibits a divergence with a critical exponent
near the critical temperature $\sim 1/(T-T_c)^\xi$,
if 3D ordering occurs through a continuous (second-order) transition.
In case of nematic regime, the nematic ordering at finite temperature 
may induce singularity in $1/T_1$.
On the other hand, at high temperature,
the effective TL-liquid theory becomes
unreliable when the energy scale $k_BT$ is increased up to the order of
$u/a$, which is of the order of the 1D couplings $J_1$ and $J_2$,
except for the saturation limit $H \to H_c$ where $u \to 0$.
Deviation from the TL-liquid theory also comes from
breaking of magnon bound states at high temperatures.
The binding energy~\cite{binding} is numerically estimated, at most, as
$E_{\rm bind} \simeq 0.39 J_2, 0.28 J_2,$ and $0.14 J_2$
for $J_1/J_2 = -2.0, -1.0$, and $-0.6$, respectively,
which gradually decreases with lowering magnetization and
vanishes at the boarder to the lower-field vector chiral phase. 
Thus, our prediction is valid in the temperature range
$J_{3D}\ll k_BT\ll J_{1D} = {\rm min}[u/a, E_{\rm bind}]$.
In the higher-temperature regime $k_BT\gg J_{1D}$,
the relaxation rate converges to a constant value.\cite{Moriya} 
We draw schematic patterns of
the temperature dependence of $1/T_1$ in the nematic TL liquid in
Figs.~\ref{fig:several_1/T_1}(a) and (b).

%%%%%%%%%%%%%%%%%%%%%%%%%%%%%%%%%%%%%%%%%%%%%%%%%%%%%%%%%%%%%
\subsection{Hyperfine coupling tensor and field direction}
Next we discuss the anisotropic case
$A_{\perp}\neq A_{\parallel}$
(we have assumed $A_{\perp:n}=A_{\parallel:n}$ so far for simplicity).
In the extreme case $A_{\perp}\gg A_{\parallel}$,
we can no longer neglect the contribution from the
exponentially decaying transverse spin
correlation in $1/T_1$. In this extreme case,
power-law behavior in $1/T_1$ of multipolar TL liquids is negligible
down to extremely low temperature, hence we observe a thermal activation form 
$\sim e^{-\Delta/(k_BT)}$ 
in $1/T_1$ in a low-temperature regime $k_BT\ll J_{1D}$.
Thus it is easy to detect a characteristic feature of multipolar TL liquids. 
If target magnets have sufficiently high crystallographic symmetry, 
the principal axes for hyperfine coupling tensor ${\cal A}_{\mu\nu}$ 
can be defined and the tensor is diagonalized. 
In this case, 
tuning the field direction parallel to a principal axis, 
we can eliminate off-diagonal elements ${\cal A}_{xz}$ and ${\cal A}_{yz}$,
and equivalently set $A_{\parallel}=0$. 
Therefore, the setup of $H$ parallel to the axis offers
an easy way of measuring the transverse spin gap and distinguishing multipolar liquids from ordinary TL liquid. 
Figures~\ref{fig:several_1/T_1}(c) and \ref{fig:several_1/T_1}(d) 
present the schematic temperature dependence of $1/T_1$ in the nematic 
and SDW$_2$ TL liquids for the anisotropic case $A_{\perp}\gg A_{\parallel}$.
In the opposite limit $A_{\perp}\ll A_{\parallel}$,
our predictions in Figs.~\ref{fig:NMR_nematic} and \ref{fig:NMR_nematic_2}
are highly reliable in $J_{3D}\ll k_BT\ll J_{1D}$. However, we should note
that in this limit $A_{\perp}\ll A_{\parallel}$,
$1/T_1$ becomes insensitive to the disappearance
of the algebraically-decaying transverse spin correlation,
the hallmark of the nematic liquid,
and is not efficient in distinguishing the usual and nematic TL liquids.

%%%%%%%%%%%%%%%%%%%%%%%%%%%%%%%%%%%%%%%%%%%%%%%%%%%%%%%%%%%%%%%%%%%%
\subsection{Effects of DM interaction}
In the rest of this section, we consider the effects of magnetic anisotropies 
which generally exist in real magnets.
In spin-$\frac{1}{2}$ systems, one of the most realistic anisotropies 
is the DM interaction defined by
\begin{align}
\label{eq:DM}
{\cal H}_{DM}(Q,\varphi,m)=
\sum_j \cos(Qj+\varphi){\bol D}\cdot({\bol S}_j\times{\bol S}_{j+m})
\end{align}
for $m=1,2$.
Possible values of $Q$, $\varphi$, and the DM vector ${\bol D}$
depend strongly on the crystal structure of each compound.
Since the DM interaction can sometimes induce an excitation gap
accompanied with local spin polarization, it may violate the multipolar
TL liquids.
We discuss the effects of the DM interaction on the nematic TL liquid
in the low-energy theory.

In the weak-coupling theory\cite{Kolezhuk,Vekua,Hikihara1}
for the spin-$\frac12$ $J_1$-$J_2$ chain with $|J_1|\ll J_2$,
the nematic phase is expressed by the effective Hamiltonian
\begin{eqnarray}
\label{eq:eff_zigzag_2}
\tilde{\cal H}_{J_1-J_2}^{\rm eff}&=&
\int dy \sum_{\gamma=\pm}\frac{v_{\gamma}}{2}
\left[K_\gamma^{-1}(\partial_y\phi_\gamma)^2
+K_\gamma(\partial_y\theta_\gamma)\right]\nonumber\\
&&
+c_1 (2a)^{-1}\sin(\sqrt{8\pi}\phi_-+\pi M)
\nonumber\\
&&+c_2 (\partial_y\theta_+)\sin(\sqrt{2\pi}\theta_-)
+\cdots
\end{eqnarray}
under the condition that only the $c_1$ term is relevant to the Gaussian model,
where $y=2ja$, $(\phi_\pm,\theta_\pm)
=(\phi_1\pm\phi_2,\theta_1\pm\theta_2)/\sqrt{2}$, and
$(\phi_{1(2)},\theta_{1(2)})$ is a boson-field pair defined
in each of the decoupled AF-$J_2$ chains (see Sec.~\ref{sec:Multipolar_2_TL}).
The coupling constants $c_{1,2}$ are proportional to $J_1$.
The $c_1$ term makes $\phi_-$ pinned.
As a result, the transverse spin correlation decays in an exponential fashion.
The remaining $(\phi_+,\theta_+)$ sector induces the gapless
behavior of longitudinal-spin and nematic correlation functions.
On the other hand, the $c_2$ term is known to induce a vector chiral
phase~\cite{Nersesyan,HikiharaKK,Kolezhuk} with
$\langle ({\bol S}_j\times{\bol S}_{j+1})^z\rangle\sim
\langle \sin(\sqrt{2\pi}\theta_-)\rangle \ne 0$ in a lower-field
regime.~\cite{Kolezhuk,Hikihara1} Let us concentrate on
the former nematic TL liquid in the following.

Using the effective theory~(\ref{eq:eff_zigzag_2}),
we investigate five typical situations:
${\cal H}_{DM}^{\rm{\rnum 1}}={\cal H}_{DM}(0,0,1)$,
${\cal H}_{DM}^{\rm{\rnum 2}}={\cal H}_{DM}(\pi,0,1)$,
${\cal H}_{DM}^{\rm{\rnum 3}}={\cal H}_{DM}(0,0,2)$,
${\cal H}_{DM}^{\rm{\rnum 4}}={\cal H}_{DM}(\pi,0,2)$, and
${\cal H}_{DM}^{\rm{\rnum 5}}={\cal H}_{DM}(\pi/2,\pi/4,2)$.
Let us first consider the case of ${\bol D}=(0,0,D_z)$, that is, 
the DM vector is parallel to the applied field $H$.
In this case, we find that the bosonized DM couplings
${\cal H}_{DM}^{{\rm\rnum 1}-{\rm\rnum 4}}$ contain slowly-moving
bosonic terms without oscillating factors $(-1)^j$ or $e^{iq\pi Mj}$ 
($q$ is an integer). They are represented as
\begin{subequations}
\label{eq:DM_boson_Dz}
\begin{eqnarray}
{\cal H}_{DM}^{\rm{\rnum 1}}&\sim& D_z\sin(\sqrt{2\pi}\theta_-)+\cdots,\\
{\cal H}_{DM}^{\rm{\rnum 2}}&\sim& D_z\Big\{\cos(\sqrt{2\pi}\theta_-)
\Big[2a\sqrt{2\pi}\partial_y\theta_-+\cdots\Big]\nonumber\\
&&-\sin(\sqrt{2\pi}\theta_-)\Big[(2a)^2\frac{\pi}{2}
\Big((\partial_y\theta_+)^2\nonumber\\
&&+(\partial_y\theta_-)^2\Big)+\cdots\Big]\Big\},\\
{\cal H}_{DM}^{\rm{\rnum 3}}&\sim& D_z\sqrt{\pi}(2a)\partial_y\theta_++\cdots,
\\
{\cal H}_{DM}^{\rm{\rnum 4}}&\sim& D_z\sqrt{\pi}(2a)\partial_y\theta_-+\cdots.
\end{eqnarray}
\end{subequations}
In general, DM terms with ${\bol D}=(0,0,D_z)$
are invariant under a global U(1) spin rotation
$S_j^+\to S_j^+ e^{i\varphi}$, which
corresponds to a shift of the phase field
$\theta_+\to\theta_+ +\sqrt{2/\pi}\varphi$. Therefore,
the bosonized DM terms with $D_z$ do not involve
any vertex operator with $\theta_+$.
In addition, all the translationally-symmetric
DM terms do not contain any vertex term with $\phi_+$ at least
for the case of incommensurate values of $M$, since
the $2n$-site translation induces $\phi_{1,2}(y)
\to \phi_{1,2}(y+2na)+n\sqrt{\pi}(1/2+M)$ and equivalently
$\phi_+(y)\to \phi_+(y+2na)+n\sqrt{\pi/2}(1+2M)$. 
Equation~(\ref{eq:DM_boson_Dz}) is consistent with these symmetry arguments.
Two DM interactions ${\cal H}_{DM}^{\rm{\rnum 1},\rm{\rnum 2}}$
tend to lock $\theta_-$ together
with the $c_2$ term in Eq.~(\ref{eq:eff_zigzag_2}), both of
which compete with the $c_1$ term.
In the nematic phase, we know that the $c_1$ term is most relevant and hence
the DM interactions ${\cal H}_{DM}^{\rm{\rnum 1},\rm{\rnum 2}}$ with 
sufficiently small $SD_z$ are negligible. In the strong DM coupling regime, 
these DM terms compete with the $c_1$ term and
could change the nematic liquid into a chiral phase with
$\langle ({\bol S}_j\times{\bol S}_{j+1})^z\rangle\neq 0$.
On the other hand, the leading terms of
${\cal H}_{DM}^{\rm{\rnum 3},\rm{\rnum 4}}$
can be absorbed into the free-boson part by shifting the field $\theta_\pm$.
The shift brings about a finite expectation value
$\langle ({\bol S}_j\times{\bol S}_{j+2})^z\rangle
\sim\langle\partial_y\theta_{1,2}\rangle$, but we note that
these chiralities do not accompany any spontaneous symmetry breaking.
The shift also affects the form of spin dynamical structure factors, 
that is, the gapless points of ${\cal S}^{\pm\mp}(k,\omega)$ are
slightly changed and a small asymmetry of $k$ dependence emerges.
In contrast to ${\cal H}_{DM}^{\rm{\rnum 1},\rm{\rnum 2}}$,
${\cal H}_{DM}^{\rm{\rnum 3},\rm{\rnum 4}}$ do not compete with
the $c_1$ term. Therefore, chirality and nematic quasi long-range order
can coexist in the latter case of ${\cal H}_{DM}^{\rm{\rnum 3},\rm{\rnum 4}}$.
This symmetry argument indicates that
the DM terms with $D_z$ do not introduce any vertex
operator in the $(\phi_+,\theta_+)$ sector, and
this statement would be true in the wide nematic-liquid region
regardless of the value of $J_1$.
The gapless nature of the $(\phi_+,\theta_+)$ sector hence survives
even after introducing these DM terms.
From these discussions, we conclude that the nematic TL liquid
survives even in the presence of DM terms with
${\bol D}=(0,0,D_z)$ at least when $D_z\ll |J_{1,2}|$.
Thus, the nematic liquid is relatively stable
against DM interactions with ${\bol D}=(0,0,D_z)$.

For the case of ${\bol D}=(D_x,0,0)$,
the U(1) rotational symmetry is broken and
the emergence of vertex operators $e^{iq\sqrt{\pi}\theta_+}$
is generally allowed. In fact, ${\cal H}_{DM}^{\rm\rnum 5}$ is bosonized as
\begin{align}
\label{eq:DM_boson_Dx}
{\cal H}_{DM}^{\rm\rnum 5}\sim D_x M\sin(\sqrt{\pi/2}\theta_+)
\cos(\sqrt{\pi/2}\theta_-)+\cdots.
\end{align}
This leading term has scaling dimension
$1/(8K_+)+1/(8K_-)$ and can generate a staggered magnetization along
the $S^y$ axis in each AF-$J_2$ chain.
In the weak $J_1$ limit, ${\cal H}_{DM}^{\rm\rnum 5}$ is more relevant than
the $c_1$ term with scaling dimension $2K_-\approx 2K$ and
it can violate the nematic liquid phase.
[In this limit, $K_\pm$ approaches the value of original
TL-liquid parameter $K$ ($1/2<K<1$) of the AF-$J_2$ chain in magnetic field.]
When $J_1$ is sufficiently strong, the $c_1$ terms defeats the perturbation
${\cal H}_{DM}^{\rm\rnum 5}$ and protects the nematic liquid phase, pinning $\phi_-$.
In this case, $\sin(\sqrt{\pi/2}\theta_+)\cos(\sqrt{\pi/2}\theta_-)$ generates
new operators $\cos(\sqrt{2\pi}\theta_+)$ and $\cos(\sqrt{2\pi}\theta_-)$
via the renormalization-group process. The first term with scaling
dimension $1/(2K_+)$ can open a gap in the $(\phi_+,\theta_+)$ sector
and induce a transverse staggered magnetization, 
since the U(1) spin symmetry is broken by the DM term.
In the weak DM coupling regime, thus, this DM term perturbatively deforms
nematic spin liquid,
opening a small gap in longitudinal spin modes
and inducing a small expectation value of transverse spins.
The other DM interactions ${\cal H}_{DM}^{\rm{\rnum 1}-\rm{\rnum 4}}$
with ${\bol D}$ parallel to the $x$ axis do not possess any 
slowly-moving operator
within a naive calculation based on the bosonization, but they
generally have the ability to generate $e^{iq\sqrt{\pi}\theta_+}$, 
which is allowed from the symmetry argument.
Thus, the gapless nature of the $(\phi_+,\theta_+)$ sector is
expected to be fragile and unstable against DM terms with
${\bol D}\neq (0,0,D_z)$.

These discussions on DM terms indicate that if we apply magnetic
field parallel to the DM vector ${\bol D}$, the nematic
TL liquid is stably realized and our prediction of $1/T_1$ is reliable
in wider $H$-$T$ space.
Even when ${\bol D}$ is not parallel to $H$, the nematic phase will 
also survive if the DM coupling is small, $|J_{1,2}| \gg |{\bol D}|$,
and if the system is in the temperature regime $k_BT \gg |{\bol D}|$.

%%%%%%%%%%%%%%%%%%%%%%%%%%%%%%%%%%%%%%%%%%%%%%%%%
%%%%%%%%%%%%%%%%%%%%%%%%%%%%%%%%%%%%%%%%%%%%%%%%%
%%%%%%%%%%%%%%%%%%%%%%%%%%%%%%%%%%%%%%%%%%%%%%%%%
%%%%%%%%%%%%%%%%%%%%%%%%%%%%%%%%%%%%%%%%%%%%%%%%%
\section{Conclusions}
\label{sec:Conclusions}

We have evaluated accurately the field and temperature dependence
of the NMR relaxation rate $1/T_1$ in magnetic quadrupolar
(spin nematic) TL liquid of the 
spin-$\frac{1}{2}$ $J_1$-$J_2$ chain, combining field-theoretical
techniques with DMRG results (see Figs.~\ref{fig:NMR_nematic}
and \ref{fig:NMR_nematic_2}).
As a comparison,
we have also calculated $1/T_1$ of the spin-$\frac{1}{2}$ AF Heisenberg chain,
using field theories, the Bethe ansatz, and DMRG method
(see Fig.~\ref{fig:NMR_usualTL}).

In the nematic and SDW$_2$ TL-liquid phase at a low temperature 
$k_BT\ll |J_{1,2}|$, 
the relaxation rate $1/T_1$ first decreases with increasing magnetic
field and then rapidly increases near saturation.
In the higher-temperature regime $k_BT\sim 0.1 J_2$, the field dependence
of $1/T_1$ becomes quite small except in the vicinity of saturation. 
The decreasing behavior of $1/T_1$ with increasing $H$ comes from the
monotonic increase in the TL-liquid parameter $\kappa$, 
while the rapid increase in $1/T_1$ near saturation is attributed to
the decrease of velocity $u\to 0$. The monotonic property of $\kappa$
is also essential for the characteristic temperature dependence$^{15}$
of $1/T_1$: 
with lowering temperature, $1/T_1$ increases in an algebraic form
in the lower-field SDW regime ($\kappa<1/2$),
whereas it decreases in the higher-field
nematic regime ($\kappa>1/2$).
These characteristic $H$ and $T$ dependences could be
a signature of the nematic and SDW$_2$ TL-liquid phase.
Similar features are
also expected to appear in higher-order multipolar TL liquids,
for example, octupolar and hexadecapolar TL liquids.
Probing the $H$ dependence would be easier than doing the $T$
dependence since the former does not necessarily require the
accession to high-field regime, where
NMR measurements are difficult. 
A combination of the $H$ and $T$ dependences of $1/T_1$ 
and the gapless behavior observed from bulk quantities 
(such as specific heat and magnetic susceptibilities) would 
present indirect but strong evidence for multipolar TL-liquid phases.

In Sec.~\ref{sec:RealMagnets},
we also considered the effects of spin anisotropies
and interchain couplings, which are
neglected in the ideal $J_1$-$J_2$ spin chain model.
In particular, we point out that
if the direction of the external field $H$ can be
parallel to the principal axis of the hyperfine-coupling tensor,
all the algebraic contributions in the temperature dependence of $1/T_1$,
that is, $A_{\parallel:n}$ terms, disappear. As a result, $1/T_1$ of
multipolar phases in the $J_1$-$J_2$ chain becomes a thermal activation
form $\sim e^{-\Delta/(k_BT)}$. In addition,
we predict that the nematic TL liquid is stable
for small DM terms with the DM vector parallel to the applied field,
while it can be easily deformed by the DM terms, thereby accompanied
with small transverse staggered magnetization, when the DM vector is
perpendicular to the field.
This indicates that we should apply magnetic field parallel
to the DM vector to obtain a stable nematic liquid phase
in real quasi-1D $J_1$-$J_2$ magnets.

Finally we comment on some real compounds. Recently, quasi-1D edge-sharing cuprate magnets, for example, 
$\rm LiCu_2O_2$, $\rm LiCuVO_4$, $\rm Rb_2Cu_2Mo_3O_{12}$ and
$\rm PbCuSO_4(OH)_2$, have been studied
extensively as low-dimensional frustrated or multiferroic systems.
Their magnetic properties are believed to be described by
spin-$\frac{1}{2}$ $J_1$-$J_2$ chains in a certain temperature regime.
Except for $\rm Rb_2Cu_2Mo_3O_{12}$, a 3D ordering has been observed
below a very low critical temperature $T_c$ at least in zero magnetic field.
Furthermore, the values of $J_1$ and $J_2$ have been
semi-quantitatively estimated in several ways. 
The estimated coupling constants and the critical temperatures in
$\rm LiCu_2O_2$ (Refs.~\onlinecite{Masuda,Masuda2,Park,Seki}),
$\rm PbCuSO_4(OH)_2$ (Refs.~\onlinecite{Kamieniarz,Baran,Yasui,Wolter}),
and $\rm Rb_2Cu_2Mo_3O_{12}$ (Ref.~\onlinecite{Hase}) are,
respectively, $(J_1/k_B,J_2/k_B,T_c)\sim(-138$~K, 86~K, 24~K), 
$\sim(-13$~K, 21~K, 2.8~K), and $(J_1/k_B, J_2/k_B)\sim(-138$~K, 51~K),
while two different results have been reported for $\rm LiCuVO_4$
(Refs.~\onlinecite{Enderle,Naito,Hagiwara,Buttgen1,Sirker}); 
$(J_1/k_B,J_2/k_B,T_c)\sim(-19$~K, 45~K, 2~K)
and $\sim(-182$~K, 91~K, 2~K). The critical temperatures $T_c$ 
are thus small compared to the magnitude of $J_{1,2}$ except for 
$\rm LiCu_2O_2$. It means that three dimensionality is small in 
these compounds.
We note that our prediction for $1/T_1$ could be applied to the temperature
condition $k_BT_c\ll k_BT\ll |J_{1,2}|$.

Among these compounds, the magnetization process of 
$\rm LiCuVO_4$ has been intensively
studied in some experimental groups.
In the low-field regime including zero field, spiral phases exist 
at low temperatures.
Above $H\approx7.5$~T, the spiral phase turns into another phase which 
was concluded by NMR measurements\cite{Buttgen1} to be a modulated collinear 
phase. The appearance of both the spiral and modulated collinear phases 
was well understood\cite{Hikihara1} as a consequence of 
the vector chiral phase and the incommensurate SDW$_2$ phase in 
the 1D $J_1$-$J_2$ spin chain. In addition, quite recently,
a new phase transition has been observed~\cite{Hagiwara}
at $H\approx 40$~T, where the saturation field is $H_{\rm s}\approx 47$~T.
Comparing the result with the phase diagram in the 1D $J_1$-$J_2$ spin
chain,\cite{Hikihara1,Sudan} we expect
the new high-field phase for $40\mbox{~T}<H<47$~T to 
be a nematic long-range ordered phase.
So far, the magnetic structure of this new phase has not
been experimentally identified at all.
We expect that NMR measurements above the critical temperatures of this
new high-field phase and of
the intermediate-field modulated collinear phase would be useful
to verify whether the new phase is a nematic phase.
We also note that $\rm PbCuSO_4(OH)_2$ has a rather small saturation field
$H_{\rm s}\approx 10$~T.~\cite{Wolter} This might be an ideal material for
measuring the $H$ dependence of $1/T_1$ up to the saturation field.

%\begin{figure}[tth]
%\begin{center}
%\includegraphics[width=8cm]{T1-All_2.eps}
%\vspace{5cm}
%\end{center}
%\caption{(color online) }
%\label{NMR_rate}
%\end{figure}

%%%%%%%%%%%%%%%%%%%%%%%%%%%%%%%%%%%%%%%%%%%%%%%%%
%%%%%%%%%%%%%%%%%%%%%%%%%%%%%%%%%%%%%%%%%%%%%%%%%
%%%%%%%%%%%        figure        %%%%%%%%%%%%%%%%
%%%%%%%%%%%%%%%%%%%%%%%%%%%%%%%%%%%%%%%%%%%%%%%%%
%%%%%%%%%%%%%%%%%%%%%%%%%%%%%%%%%%%%%%%%%%%%%%%%%
%%%%%%%%%%%%%%%%%%%%%%%%%%%%%%%%%%%%%%%%%%%%%%%%%
%\begin{figure}[b]
%\end{figure}

%%%%%%%%%%%%%%%%%%%%%%%%%%%%%%%%%%%%%%%%%%%%%%%%%%%%%%%%%
%%%%%%%%%%%%%%%%%%%%%%%%%%%%%%%%%%%%%%%%%%%%%%%%%%%%%%%%%
%%%%%%%%%%%%%%%%%%%%%%%%%%%%%%%%%%%%%%%%%%%%%%%%%%%%%%%%%
%%%%%%%%%%%%%%%%%%%%%%%%%%%%%%%%%%%%%%%%%%%%%%%%%%%%%%%%%
%%%%%%%%%%%%%%%%%%%%%%%%%%%%%%%%%%%%%%%%%%%%%%%%%%%%%%%%%
%%%%%%%%%%%%%%%%%%%%%%%%%%%%%%%%%%%%%%%%%%%%%%%%%%%%%%%%%

\begin{acknowledgements}
We thank Akira Furusaki, Masayuki Hagiwara, Tetsuaki Itou, Anja Wolter, 
and Yukio Yasui for fruitful discussions.
This work was supported by Grants-in-Aid for Scientific Research
from MEXT, Japan (Grants No.\ 21740277, No.\ 21740295, No.\ 22014016).
\end{acknowledgements}


\begin{thebibliography}{}

%%%%%%%%%%%%%%%%%%%%%%%%%%%%%%%%%%%%%%%%%%%%%%%%%
%%%%%%  Sec.1 Introduction   %%%%%%%%%%%%%%%%%%%%
%%%%%%%%%%%%%%%%%%%%%%%%%%%%%%%%%%%%%%%%%%%%%%%%%

\bibitem{Andreev}
A. F. Andreev and I. A. Grishchuk, Sov. Phys. JETP \textbf{60}, 267 (1984).

\bibitem{Wen}
X. G. Wen, F. Wilczek, and A. Zee, Phys. Rev. B \textbf{39}, 11413 (1989).

%%%%%%%%%%%%%  nematic  %%%%%%%%%%%%%%%%%%%
\bibitem{Chubukov}
A.~V.~Chubukov, Phys.\ Rev.\ B {\bf 44}, 4693 (1991).

\bibitem{Momoi0} T. Momoi and N. Shannon, Prog. Theor. Phys. Suppl. \textbf{159}, 72 (2005).

\bibitem{Momoi1}N.~Shannon, T.~Momoi, and P.~Sindzingre,
Phys.\ Rev.\ Lett.\ {\bf 96}, 027213 (2006).

\bibitem{Momoi2}
T.~Momoi, P.~Sindzingre, and N.~Shannon,
Phys.\ Rev.\ Lett.\ {\bf 97}, 257204 (2006).

\bibitem{Kecke}
L.~Kecke, T.~Momoi, and A.~Furusaki, Phys.\ Rev.\ B {\bf 76},
060407(R) (2007).

\bibitem{Hikihara1}
T.~Hikihara, L.~Kecke, T.~Momoi, and A.~Furusaki,
Phys.\ Rev.\ B \textbf{78}, 144404 (2008).

\bibitem{ShindouM}
R.\ Shindou and T.\ Momoi, Phys.\ Rev.\ B \textbf{80}, 064410 (2009).

\bibitem{Vekua}
T.~Vekua, A.~Honecker, H.-J.~Mikeska, and F.~Heidrich-Meisner,
Phys.\ Rev.\ B {\bf 76}, 174420 (2007).

\bibitem{Sudan}
J.~Sudan, A.~Luscher, and A.M.~L\"auchli,
Phys.\ Rev.\ B {\bf 80}, 140402(R) (2009).

\bibitem{Ueda}
H. T. Ueda and K. Totsuka, Phys. Rev. B \textbf{80}, 014417 (2009).

\bibitem{Nishimoto}
S. Nishimoto, S.-L. Drechsler, R.O. Kuzian, J. van den Brink,
J. Richter, W.E.A. Lorenz, Y. Skourski, R. Klingeler, and
B. Buechner, arXiv:1004.3300.

\bibitem{Zhitomirsky}
M. E. Zhitomirsky and H. Tsunetsugu, Europhys. Lett. {\bf 92}, 37001
(2010).
%arXiv:1003.4096.


%%%%%%%%%%%%%%  multipolar in zigzag chains  %%%%%%%%%%%%%%%

\bibitem{Sato09}
M.~Sato, T.~Momoi and A.~Furusaki,
Phys.\ Rev.\ B \textbf{79}, 060406(R) (2009).



%%%%%%%%%  materials    %%%%%%%%%%%%%%%%%


%%%%%%%%  LiCuVO4   %%%%%%%%%%%%
\bibitem{Enderle}
M. Enderle, C. Mukherjee, B. F\r{a}k, R. K. Kremer, J.-M. Broto, H. Rosner,
S.-L. Drechsler, J. Richter, J. Malek, A. Prokofiev, W. Assmus,
S. Pujol, J.-L. Raggazzoni, H. Rakoto, M. Rheinst\"adter, and
H. M. R\o nnow, Europhys. Lett. {\bf 70}, 237 (2005).

\bibitem{Hagiwara}
L. E. Svistov, T. Fujita, H. Yamaguchi, S. Kimura, K. Omura,
A. Prokofiev, A. I. Smirnov, Z. Honda and M. Hagiwara, 
J.\ Exp.\ Theo.\ Phys.\ Lett.\ {\bf 93}, 24 (2011); arXiv:1005.5668.

\bibitem{Naito}
Y. Naito, K. Sato, Y. Yasui, Y. Kobayashi, Y. Kobayashi, and M. Sato,
J. Phys. Soc. Jpn. {\bf 76}, 023708 (2007).
%M.~Enderle {\it et al}., Europhys.\ Lett.\ {\bf 70}, 237 (2005);
%Y.~Naito {\it et al}., J.\ Phys.\ Soc.\ Jpn.\ {\bf 76}, 023708 (2007).


\bibitem{Buttgen1} %%%%%%%  NMR study of collinear SDW phase   %%%%%%%%
N. B\"uttgen, H.-A. Krug von Nidda, L. E. Svistov, L. A. Prozorova,
A. Prokofiev, and W. Assmus, Phys. Rev. B \textbf{76}, 014440 (2007).



%%%%%%%%%%%%%%  Rb2Cu2Mo3O12   %%%%%%%%%%%%%
\bibitem{Hase}
M. Hase, H. Kuroe, K. Ozawa, O. Suzuki, H. Kitazawa, G. Kido, and
T. Sekine, Phys. Rev. B {\bf 70}, 104426 (2004).
%M.~Hase {\it et al}., Phys.\ Rev.\ B {\bf 70}, 104426 (2004).



%%%%%%  PbCuSO4OH2  %%%%%%%%%%%
\bibitem{Kamieniarz}
G. Kamieniarz, M. Bieli\'nski, G. Szukowski, R. Szymczak, S. Dyeyev,
and J. -P. Renard, Comp. Phys. Comm. {\bf 147}, 716 (2002).

\bibitem{Baran}
M. Baran, A. Jedrzejczak, H. Szymczak, V. Maltsev, G. Kamieniarz,
G. Szukowski, C. Loison, A. Ormeci, S.-L. Drechsler, and H. Rosner,
Phys. Stat. Sol. (c) {\bf 3}, 220 (2006).

\bibitem{Yasui}
Y. Yasui {\it et. al.}, submitted to J. Phys. Soc. Jpn. (unpublished). 

\bibitem{Wolter}
A.U.B. Wolter, R. Vogel, Y. Skourski, K.C. Rule, S. S\"ullow,
G. Heide, and B. Buechner,
International conference 
"Magnetic resonance in highly frustrated magnetic systems",
Slovenia, Feb. 2010.


%%%%%% LiCu2O2   %%%%%%%%
\bibitem{Masuda}
T. Masuda, A. Zheludev, A. Bush, M. Markina, and A. Vasiliev,
Phys. Rev. Lett. {\bf 92}, 177201 (2004).

\bibitem{Masuda2}
T. Masuda, A. Zheludev, B. Roessli, A. Bush, M. Markina, and A. Vasiliev,
Phys. Rev. B {\bf 72}, 014405 (2005).


\bibitem{Park}
S. Park, Y. J. Choi, C. L. Zhang, and S-W. Cheong, Phys. Rev. Lett.
{\bf 98}, 057601 (2007).

\bibitem{Seki}
S. Seki, Y. Yamasaki, M. Soda, M. Matsuura,
K. Hirota, and Y. Tokura, Phys. Rev. Lett. {\bf 100}, 127201 (2008).
%T.~Masuda {\it et al}., Phys.\ Rev.\ Lett.\ {\bf 92}, 177201 (2004);
%S.~Park {\it et al}., Phys.\ Rev.\ Lett.\ {\bf 98}, 057601 (2007);
%S.~Seki {\it et al}., Phys.\ Rev.\ Lett.\ {\bf 100}, 127201 (2008).


%%%%%%%%% NaCu2O2    %%%%%%%%%%%%%%
\bibitem{Drechsler}
S.-L. Drechsler, J. Richter, A. A. Gippius, A. Vasiliev, A. A. Bush,
A. S. Moskvin, J. M\'alek, Yu. Prots, W. Schnelle, and H. Rosner,
Europhys. Lett. {\bf 73}, 83 (2006).
%S.-L.~Drechsler {\it et al}., Europhys.\ Lett.\ {\bf 73}, 83 (2006).



%%%%%%%%%%%%%%%%%%%%%%%%%%%%%%%%%%%%%%%%%%%%%%%%%
%%%%%%      Sec.2  NMR and 1/T_1      %%%%%%%%%%%
%%%%%%%%%%%%%%%%%%%%%%%%%%%%%%%%%%%%%%%%%%%%%%%%%

%%%%%%%%%%%%%%%%  comment on 1/T_1   %%%%%%%%%%%%%%%
\bibitem{Slichter}
C. P. Slichter, {\it Principles of Magnetic Resonance}
(Springer-Verlag, Berlin, 1990).


\bibitem{Goto}
T.~Goto, T.~Ishikawa, Y.~Shimaoka, and Y.~Fujii, Phys.\ Rev.\ B {\bf 73},
214406 (2006).




%%%%%%%%%%%%%%%%%%%%%%%%%%%%%%%%%%%%%%%%%%%%%%%%%
%%%%%%  Sec.3 Multipolar and usual TL liquids %%%
%%%%%%%%%%%%%%%%%%%%%%%%%%%%%%%%%%%%%%%%%%%%%%%%%

\bibitem{incommeNematic}
In the narrow range $-2.720<J_1/J_2<-2.669$,
the soft two-magnon bound state has an incommensurate momentum
$k\ne\pi$ to realize a two-component TL liquid.
In this paper, we do not consider this case and focus on
the one-component mulitpolar TL liquids of the multimagnon bound states
with momentum $k=\pi$. However, it is naively expected that 
the incommensurate multipolar liquid qualitatively shares the features 
of $1/T_1$ in the commensurate multipolar liquids 
(see sections III and IV) which mainly result from 
the formation of multimagnon bound states.  


%%%%%%%%%  Haldane's bosonization   %%%%%%%%%%%%%%%%%
\bibitem{Haldane}
F. D. M. Haldane, Phys.\ Rev.\ Lett.\ {\bf 47}, 1840 (1981).

%%%%%%%%%%%%%%%%%%%  text CFT and bosonization   %%%%%%%%%%%%%%%%%%%
\bibitem{Gia_text}
T.~Giamarchi, {\it Quantum Physics in One Dimension}
(Oxford University Press, New York, 2004).


%%%%%%%%%  basis of NMR in spin chains   %%%%%%%%%%%%%%%%%
\bibitem{Giamarchi}
R.~Chitra and T.~Giamarchi, Phys.\ Rev.\ B {\bf 55}, 5816 (1997).

\bibitem{Giamarchi2}
T. Giamarchi and A. M. Tsvelik, Phys.\ Rev.\ B {\bf 59}, 11398 (1999).

\bibitem{Giamarchi3}
M. Klanj\u sek, H. Mayaffre, C. Berthier, M. Horvati\'c, 
B. Chiari, O. Piovesana, P. Bouillot, C. Kollath, E. Orignac,
R. Citro, and T. Giamarchi, Phys. Rev. Lett. {\bf 101}, 137207 (2008). 

\bibitem{Giamarchi4}
P. Bouillot, C. Kollath, A. M. L\"auchli, M. Zvonarev, 
B. Thielemann, C. R\"uegg, E. Orignac, R. Citro, M. Klanj\u sek, 
C. Berthier, M. Horvati\'c, and T. Giamarchi, arXiv:1009.0840.

\bibitem{BogoIK1986}
N.\ M.\ Bogoliubov, A.\ G.\ Izergin, and V.\ E.\ Korepin,
Nucl.\ Phys.\ B {\bf 275}, 687 (1986).

\bibitem{QinFYOA1997}
S.\ Qin, M.\ Fabrizio, L.\ Yu, M.\ Oshikawa, and I.\ Affleck,
Phys.\ Rev.\ B {\bf 56}, 9766 (1997).

\bibitem{CabraHP1998}
D.\ C.\ Cabra, A.\ Honecker, and P.\ Pujol,
Phys.\ Rev.\ B {\bf 58}, 6241 (1998).

\bibitem{Essler}
For instance, the list of the values of $K$ and $v$ are summarized in
F. H. L. Essler, A. Furusaki, and T. Hikihara,
Phys. Rev. B {\bf 68}, 064410 (2003).

\bibitem{Hikihara2}
T.~Hikihara, and A.~Furusaki, Phys. Rev. B {\bf 69}, 064427 (2004);
The detailed list of $a_1$ and $b_0$ is presented in arXiv:cond-mat/0310391.


%%%%%%%%%%%%%%%%%  NMR  theory and exp near SU(2) point   %%%%%%%%%%%%%%
\bibitem{Sachdev}
There are logarithmic corrections 
to Eq.~(\ref{eq:1/T_1_TL}) near the zero-field SU(2)-symmetric point
with $K=1/2$, which we omitted. See
S. Sachdev, Phys.\ Rev.\ B \textbf{50}, 13006 (1994);
M.~Takigawa, N.~Motoyama, H.~Eisaki, and S.~Uchida,
Phys. Rev. Lett. {\bf 76}, 4612 (1996); M.~Takigawa, O.A.~Starykh,
A.W.~Sandvik, and R.R.P.~Singh, Phys. Rev. B {\bf 56}, 13681 (1997).
%M.~Takigawa {\it et al}., Phys.\ Rev.\ Lett.\ {\bf 76}, 4612 (1996);
%M.~Takigawa {\it et al}., Phys.\ Rev.\ B {\bf 56}, 13681 (1997).


%%%%%%%%%%%%%%%%%%%%%%%%%%%%%%%%%%%%%%%%%%%%%%%%%
%%%%%%  Sec.4  Field and Temp dep 1/T_1  %%%%%%%%
%%%%%%%%%%%%%%%%%%%%%%%%%%%%%%%%%%%%%%%%%%%%%%%%%


\bibitem{Hikihara3}
T.~Hikihara, T.~Momoi, A.~Furusaki, and H.~Kawamura,
Phys.\ Rev.\ B {\bf 81}, 224433 (2010).
%arXiv:1004.0622.


%%%%%%%%% 1/T_1 in high-field region in S=1/2 Heisenberg chain %%%%
\bibitem{Azevedo}
L. J. Azevedo, A. Narath, Peter M. Richards, and Z. G. Soos,
Phys.\ Rev.\ Lett.\ {\bf 43}, 875 (1979).
\bibitem{Azevedo2}
L. J. Azevedo, A. Narath, P. M. Richards, and Z. G. Soos,
Phys.\ Rev.\ B {\bf 21}, 2871 (1980).

\bibitem{Groen}
J. P. Groen, T. O. Klaassen, N. J. Poulis, G. M\"uller, H. Thomas,
and H. Beck, Phys.\ Rev.\ B {\bf 22}, 5369 (1980).

\bibitem{Kuhne}
H. K\"uhne, M. G\"unther, S. Grossjohann, W. Brenig, F. J. Litterst,
A. P. Reyes, P. L. Kuhns, M. M. Turnbull, C. P. Landee, and H.-H. Klauss,
Phys. Status Solidi B {\bf 247}, 671 (2010).

%\bibitem{Zvyagin}
%H. K\"uhne,1 A.A. Zvyagin, M. G\"unther, A.P. Reyes,
%P.L. Kuhns, M.M. Turnbull, C.P. Landee, and H.-H. Klauss, arXiv:1007.3608.



%%%%%%%%% 1/T_1 in high-field region in S=5/2 Heisenberg chain %%%%
%\bibitem{Boucher}
%J. P. Boucher and J. P. Renard,
%Phys.\ Rev.\ Lett.\ {\bf 45}, 486 (1980).




%%%%%%%%%%%%%%  zigzag AF-J1  %%%%%%%%%%%%%%%
\bibitem{Okunishi}
K.~Okunishi and T.~Tonegawa, J.\ Phys.\ Soc.\ Jpn.\ {\bf 72}, 479
(2003): In this paper, the nematic TL liquid is called the even-odd phase.


%%%%%%%%%%%%%%  magnon binding energy  %%%%%%%%%%%%%%%
\bibitem{binding}
The magnon binding energy is defined as follows.
First we prepare $\Delta_n({\cal M})=E_0({\cal M}+n)-E_0({\cal M})$, where
$E_0({\cal M})$ is the lowest energy eigenvalue in the subspace
with fixed $\sum_j S_j^z={\cal M}=ML$
($L$ is the total number of sites). 
Calculating the binding energies of the two definitions 
$E_{+}=2\Delta_1-\Delta_2$ and $E_{-}=2\Delta_{-1}-\Delta_{-2}$, 
we obtain the true binding energy from the minimum value of them,
$E_{\rm bind}={\rm min}(E_{+},E_{-})$. 


%%%%%%%%%%%%%%%%%%%%%%%%%%%%%%%%%%%%%%%%%%%%%%%%%
%%%%%%  Sec.5  Discussion on real magnets  %%%%%%
%%%%%%%%%%%%%%%%%%%%%%%%%%%%%%%%%%%%%%%%%%%%%%%%%

\bibitem{Moriya}  T.\ Moriya, Prog.\ Theor.\ Phys.\ \textbf{16}, 23 (1956);
T.\ Moriya, Prog.\ Theor.\ Phys.\ \textbf{16}, 641 (1956).

%%%%%%%%%%%%%%%%%%%  vector chiral phase in zigzag chains   %%%%%%%%%%%
\bibitem{Kolezhuk}
A.~Kolezhuk and T.~Vekua, Phys.\ Rev.\ B {\bf 72}, 094424 (2005).

\bibitem{Nersesyan}
A.A.~Nersesyan, A.O.~Gogolin, and F.H.L.~E\ss ler, Phys.\ Rev.\ Lett.
{\bf 81}, 910 (1998).

\bibitem{HikiharaKK}
T.\ Hikihara, M.\ Kaburagi, and H.\ Kawamura,
Phys.\ Rev.\ B, {\bf 63}, 174430 (2001).

%%%%%%%%%%%%%%%%%%%%%%%%%%%%%%%%%%%%%%%%%%%%%%%%%
%%%%%%  Sec.6  Conclusions        %%%%%%%%%%%%%%%
%%%%%%%%%%%%%%%%%%%%%%%%%%%%%%%%%%%%%%%%%%%%%%%%%


%%%%%%%%%  Theoretical esitation J_{1,2} of several compounds %%%%%%%%
\bibitem{Sirker}
J. Sirker, Phys. Rev. B{\bf 81}, 014419 (2010).










\end{thebibliography}
\end{document}